\documentclass[11pt, a4paper, logo, copyright]{googlecloud}

\pdftrailerid{redacted}

\makeatletter
\renewcommand\bibentry[1]{\nocite{#1}{\frenchspacing\@nameuse{BR@r@#1\@extra@b@citeb}}}
\makeatother

\usepackage{amsmath,amsfonts,bm}

\def\eqref#1{equation~\ref{#1}}
\def\1{\bm{1}}

\DeclareMathAlphabet{\mathsfit}{\encodingdefault}{\sfdefault}{m}{sl}
\SetMathAlphabet{\mathsfit}{bold}{\encodingdefault}{\sfdefault}{bx}{n}

\RequirePackage{algorithm}
\RequirePackage{algorithmic}

\usepackage[symbol]{footmisc}
\usepackage{natbib}
\usepackage{tcolorbox}
\usepackage{times}
\usepackage[T1]{fontenc}
\usepackage{textcomp}
\usepackage{upquote}
\usepackage{latexsym}
\usepackage{kantlipsum, lipsum}
\usepackage{dsfont}
\usepackage{makecell}
\usepackage{url}            % simple URL typesetting
\usepackage{nicefrac}       % compact symbols for 1/2, etc.
\usepackage{multicol}
\usepackage{bbm}
\usepackage{multirow}
\usepackage{adjustbox}
\usepackage{listings}
\usepackage{soul}
\usepackage{float}
\usepackage{wrapfig}
\usepackage{needspace}
\usepackage{blindtext}
\usepackage{tablefootnote}
\usepackage{amsfonts}
\usepackage[flushleft]{threeparttable}
\usepackage{bbding}
\usepackage{xcolor}
\usepackage{xspace}
\usepackage{bm}
\usepackage{arydshln}
\usepackage{enumitem}
\usepackage{setspace}
\usepackage{color}
\usepackage{longtable}
\usepackage[normalem]{ulem}
\usepackage{ulem}
\usepackage[nomargin,inline,marginclue,draft]{fixme}
\usepackage{balance}
\usepackage{verbatim}
\usepackage{diagbox}
\usepackage{changepage}
\usepackage{pifont}
\usepackage{array}   % Optional: Improves table column alignment

\usepackage{microtype}
\usepackage{graphicx}
\usepackage{fontawesome5}
\usepackage{twemojis}
\usepackage{subcaption}
\usepackage{booktabs} % for professional tables

\usepackage{titletoc} % Automatically generated appendix-only contents.
\usepackage{hyperref}

\usepackage{amsmath}
\usepackage{amssymb}
\usepackage{mathtools}
\usepackage{amsthm}

\usepackage[capitalize,noabbrev]{cleveref}

\theoremstyle{plain}

\theoremstyle{definition}

\theoremstyle{remark}

\usepackage[textsize=tiny]{todonotes}
\usepackage{xcolor}

\usepackage{pifont}
\newcommand{\cmark}{\textcolor{blue}{\ding{51}}}  % tick (yes)
\newcommand{\xmark}{\textcolor{red}{\ding{55}}}  % cross (no)

\let\cite\citep

\usepackage[table]{xcolor}

\tcbuselibrary{skins,breakable}
\newtcolorbox{promptbox}[1]{
  colback=white,
  colframe=blue!100,
  colbacktitle=blue!10,
  coltitle=black,
  title=#1,
  fonttitle=\bfseries\sffamily,
  breakable,
  enhanced,
  boxrule=0.8pt
}

\usepackage{listings}

\lstdefinestyle{promptblock}{
  basicstyle=\ttfamily\footnotesize,
  columns=fullflexible,
  keepspaces=true,
  breaklines=true,
  breakatwhitespace=false,
  showstringspaces=false,
  frame=single,
  framerule=0.3pt,
  rulecolor=\color{black!25},
  backgroundcolor=\color{black!2},
  captionpos=b,
  aboveskip=6pt,
  belowskip=6pt
}

\newcommand{\promptlisting}[3]{%
  \IfFileExists{#3}{%
    \lstinputlisting[style=promptblock,caption={#1},label={#2}]{#3}%
  }{%
    \refstepcounter{lstlisting}%
    \label{#2}%
    \noindent\fcolorbox{black!25}{black!2}{%
      \parbox{0.97\linewidth}{\ttfamily\footnotesize [Prompt artifact unavailable in this workspace snapshot.]}
    }\par
    \smallskip
    \textbf{Listing~\thelstlisting.} #1\par\medskip
  }%
}
\newcommand{\methodname}{\textsc{T2J-Bench}\xspace}
\newcommand{\harnessname}{\textsc{TwinHarness}\xspace}

\title{\methodname: Benchmarking Coding Agents on Execution-Grounded PyTorch-to-JAX Repository Migration}

\correspondingauthor{taofeng2@illinois.edu, jiefengc@google.com, jinsungyoon@google.com}

\author[2]{Tao Feng$^\dagger$}
\author[1]{Jiefeng Chen$^\dagger$}
\author[3]{Linxin Song}
\author[1]{Yufan Zhuang}
\author[4]{Yue Huang}
\author[1]{Bhavana Dalvi Mishra}
\author[5]{Chi Wang}
\author[1]{Xiangru Tang}
\author[3]{Jieyu Zhao}
\author[2]{Ge Liu}
\author[1]{Tomas Pfister}
\author[1]{Jinsung Yoon}

\affil[1]{Google Cloud AI Research}
\affil[2]{University of Illinois at Urbana-Champaign}
\affil[3]{University of Southern California}
\affil[4]{University of Notre Dame}
\affil[5]{Google Deepmind}

\begin{abstract}
Migrating a machine-learning codebase across frameworks---for example from PyTorch to
JAX---is a routine but costly engineering task that is fundamentally \emph{cross-paradigm}:
eager, stateful, object-oriented modules must be rewritten as pure, jittable functions over
an explicit parameter pytree. We ask whether coding agents can perform such
whole-repository migrations \emph{correctly}, and argue that existing code benchmarks,
which grade unit-test pass rates or surface-level similarity to a reference translation,
cannot answer this: they neither require the migrated system to \emph{behave} the same nor
prevent a JAX reimplementation from faking equivalence by secretly importing PyTorch itself. 
We introduce \methodname, a benchmark of 63 whole-repository PyTorch$\to$JAX
migration tasks spanning 18 model families across both training and serving, where
torch-free checkpoint conversion is itself part of the deliverable. \methodname\ grades
\emph{execution-grounded functional equivalence}: each candidate is \emph{run} inside a
separate, \texttt{torch}-free, offline verifier and scored against references frozen from
the PyTorch original, under a hard independence gate and a continuous, threshold-free
reward. Across twenty off-the-shelf agent--model configurations, the strongest
recovers only $0.676$ of the reward against a $0.980$ oracle, with
substantial numeric and behavioral divergence despite high spec pass rates.
Finally, we contribute \harnessname, a reference-grounded agent harness that turns the
agent's executable PyTorch source into legitimate self-verification feedback; applied to an
\texttt{opus-5} agent it raises the reward from $0.649$ to $0.723$---the highest of any
configuration---using only task-visible resources. 

\end{abstract}

\begin{document}

\maketitle

\begin{abstract}
Migrating a machine-learning codebase across frameworks---for example from PyTorch to
JAX---is a routine but costly engineering task that is fundamentally \emph{cross-paradigm}:
eager, stateful, object-oriented modules must be rewritten as pure, jittable functions over
an explicit parameter pytree. We ask whether coding agents can perform such
whole-repository migrations \emph{correctly}, and argue that existing code benchmarks,
which grade unit-test pass rates or surface-level similarity to a reference translation,
cannot answer this: they neither require the migrated system to \emph{behave} the same nor
prevent a JAX reimplementation from faking equivalence by secretly importing PyTorch itself. 
We introduce \methodname, a benchmark of 63 whole-repository PyTorch$\to$JAX
migration tasks spanning 18 model families across both training and serving, where
torch-free checkpoint conversion is itself part of the deliverable. \methodname\ grades
\emph{execution-grounded functional equivalence}: each candidate is \emph{run} inside a
separate, \texttt{torch}-free, offline verifier and scored against references frozen from
the PyTorch original, under a hard independence gate and a continuous, threshold-free
reward. Across twenty off-the-shelf agent--model configurations, the strongest
recovers only $0.676$ of the reward against a $0.980$ oracle, with
substantial numeric and behavioral divergence despite high spec pass rates.
Finally, we contribute \harnessname, a reference-grounded agent harness that turns the
agent's executable PyTorch source into legitimate self-verification feedback; applied to an
\texttt{opus-5} agent it raises the reward from $0.649$ to $0.723$---the highest of any
configuration---using only task-visible resources. 

% We will release the benchmark to support reproducible study of cross-paradigm ML migration.
\end{abstract}

\begin{center}
  \begingroup
  \hypersetup{hidelinks}
  \color{blue!50!black}\large
  \href{https://ft2023.github.io/T2j-web/}{\faGlobe\enspace\texttt{Website}}
  \qquad
  \href{https://github.com/google-research/t2j-bench/}{\faGithub\enspace\texttt{Code}}
  \qquad
  \href{https://huggingface.co/buckets/jiefengcgg/t2j-bench}{\texttwemoji{1f917}\enspace\texttt{Hugging Face}}
  \endgroup
\end{center}

\section{Introduction}
\label{sec:intro}

Moving model repositories across the fragmented machine-learning ecosystem is routine but costly.
Teams migrate PyTorch \citep{paszke2019pytorch} codebases to JAX
\citep{jax2018github,flax2020github} to reach new accelerators, reuse published models in
different research stacks, or build functional serving engines. This \emph{whole-repository}
job spans model definitions, training or serving loops, data pipelines, and checkpoint
formats. Repository-editing coding agents are plausible tools for this migration, but can
they produce a JAX reimplementation that both runs and \emph{behaves the same} as the original?

\begin{figure}[!t]
  \centering
  \includegraphics[width=\linewidth]{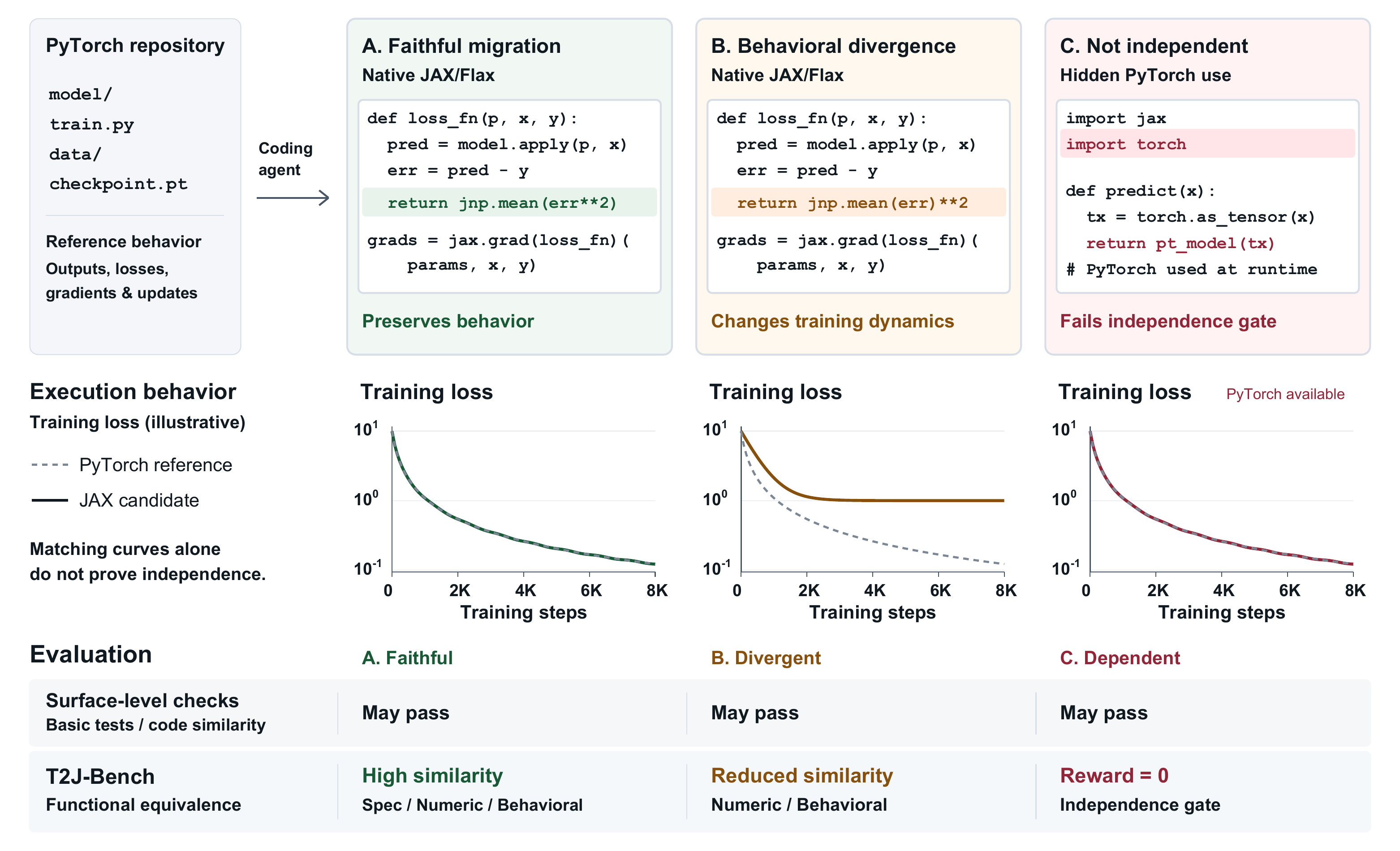}
  \caption{Why execution-grounded evaluation matters. The same PyTorch repository can yield a faithful JAX migration (A), a native JAX implementation with divergent training behavior (B), or a candidate that still depends on PyTorch (C). Surface-level checks may not distinguish these outcomes. \methodname evaluates \textsc{Spec}, \textsc{Numeric}, and \textsc{Behavioral} fidelity against frozen PyTorch references in an offline, \texttt{torch}-free verifier, with a hard independence gate. Code and loss curves are illustrative, not measured results; C's apparent agreement assumes PyTorch is available outside the verifier.}
  \label{fig:migration-motivation}
\end{figure}

General-purpose code benchmarks do not fully capture this setting. They grade functional correctness through
unit-test pass rates \citep{jimenez2024swe,chen2021evaluating} or, for transpilation,
through surface-level similarity to a reference translation
\citep{lu2021codexglue,ren2020codebleu}. Neither captures two key difficulties. First, migration is \emph{cross-paradigm}: eager, stateful, object-oriented PyTorch modules must be
re-expressed as pure, jittable functions over an explicit parameter pytree in
\texttt{flax.linen}/\texttt{optax} or \texttt{flax.nnx} --- a structural rewrite of the
program's execution model, not a line-by-line transliteration. Second, ``correct'' means
the migrated system reproduces the original's \emph{execution} --- its forward numerics,
its gradients, its optimizer trajectory, the generalization of a model it trains, and the
decode behavior of a serving engine --- to floating-point tolerances, properties that aggregate test pass rates alone do not characterize. A JAX reimplementation can also fake equivalence by importing PyTorch, so evaluation must establish \emph{independence}.
Figure~\ref{fig:migration-motivation} illustrates the distinction: a plausible native JAX migration can alter training dynamics, while a PyTorch-dependent candidate can appear numerically faithful. Evaluating migration therefore requires both execution-grounded behavioral comparison and explicit source-framework independence.

Recent work directly addresses adjacent settings. \citet{liang2026agentic}
combine in-context examples, source-execution oracles, and self-debugging for
PyTorch-to-JAX migration, including human-assessed repository conversions.
JAXBench \citep{tschand2026jaxbench} evaluates correctness and speedup of
TPU kernels generated from JAX workloads. These efforts motivate a complementary
benchmark of automatically graded whole-repository migration, including training
and serving behavior under an explicit source-framework independence gate.
Table~\ref{tab:benchmark-comparison} summarizes these distinctions.

\begin{table*}[t]
\centering
\small
\caption{Representative benchmarks and systems.
\emph{Whole-repo}: full-repository tasks; \emph{Cross-parad.}: eager/stateful-to-functional
migration; \emph{Exec-ground.}: execution-based evaluation; \emph{Behav.\ equiv.}:
reference-based training or serving behavior beyond isolated outputs; \emph{Indep.\ gate}:
source-framework exclusion during grading. A cross marks a criterion not targeted or
explicitly enforced. \citet{liang2026agentic} and JAXBench \citep{tschand2026jaxbench}
evaluate output-level equivalence without systematic trajectory or serving-state
evaluation; \citeauthor{liang2026agentic}'s repository results are human-assessed.}
\vspace{-3pt}
\label{tab:benchmark-comparison}
\adjustbox{max width=\linewidth}{%
\begin{tabular}{@{}lccccc@{}}
\toprule
Benchmark / system & \makecell{Whole-\\repo} & \makecell{Cross-\\parad.} & \makecell{Exec-\\ground.} & \makecell{Behav.\\equiv.} & \makecell{Indep.\\gate} \\
\midrule
HumanEval \citep{chen2021evaluating}              & \xmark & \xmark & \cmark & \xmark & \xmark \\
CodeXGLUE \citep{lu2021codexglue,ren2020codebleu} & \xmark & \xmark & \xmark & \xmark & \xmark \\
SWE-bench \citep{jimenez2023swebench}             & \cmark & \xmark & \cmark & \xmark & \xmark \\
MigrationBench \citep{liu2025migrationbench,wang2024repotransbench} & \cmark & \xmark & \cmark & \xmark & \xmark \\
Agentic migration \citep{liang2026agentic} & \cmark & \cmark & \cmark & \xmark & \xmark \\
JAXBench \citep{tschand2026jaxbench} & \xmark & \xmark & \cmark & \xmark & \xmark \\
\midrule
\rowcolor{gray!15}
\textbf{\methodname} (ours) & \cmark & \cmark & \cmark & \cmark & \cmark \\
\bottomrule
\end{tabular}%
}
\vspace{-3pt}
\end{table*}

We introduce \methodname: 63 whole-repository tasks requiring coding agents to convert real,
self-contained PyTorch ML repositories into native, \texttt{torch}-free JAX/Flax repositories
that preserve functional behavior
(\S\ref{sec:benchmark}). Tasks cover training migration (SFT, DPO, and PPO across 15 model families) and serving
migration (vLLM-style paged-attention engines), including weight conversion.
Grading is based on \emph{execution-grounded functional equivalence}
(\S\ref{sec:bench-grading}): every candidate is \emph{run} inside a separate verifier with
\texttt{torch} removed and network access disabled, and compared against references recorded
from the original. Independence is a hard multiplicative gate --- a JAX reimplementation that
relies on PyTorch earns nothing --- while numeric and behavioral parity are scored with
continuous, threshold-free similarity measures. Thus, \methodname rewards equivalent \emph{behavior}, not surface resemblance.

Our baseline study shows this setting is far from saturated (\S\ref{sec:experiments}):
across twenty baseline agent--model configurations evaluated with three random seeds over the full suite, the
strongest off-the-shelf configuration reaches a reward of only $0.676$ against a $0.980$
faithful JAX reimplementation ceiling --- a gap of $0.304$. High spec pass rates do not ensure numerical or behavioral fidelity:
strong agents can produce structurally valid repositories whose forward numerics,
gradients, or training dynamics diverge from the original. The reward--cost
analysis further shows that continued baseline refinement remains below
\harnessname at similar plotted API expenditure (\S\ref{sec:exp-reward-cost}). These results identify a deficiency: standard harnesses rarely turn the executable source
implementation into structured feedback on behavioral preservation. Our \emph{reference-grounded
migration harness}, \harnessname, treats the task-visible PyTorch repository as an executable
specification within an iterative \emph{diagnose--migrate--verify} loop (\S\ref{sec:harness}):
reference-grounded diagnostics localize discrepancies before further edits, while an explicit
information boundary keeps the harness from accessing hidden tests, frozen references,
verifier code, or benchmark rewards --- converting \emph{reference availability} into
\emph{reference-grounded feedback} without weakening the evaluation contract.

This additional feedback substantially improves migration quality. With the underlying
agent--model configuration held fixed, \harnessname raises the reward from $0.649$ to
$0.723$, an $11.4\%$ relative improvement over that off-the-shelf scaffold and the highest
reward of any configuration (\S\ref{sec:experiments}). These gains suggest that a meaningful part of the remaining difficulty concerns how
systematically agents use the available executable behavioral specification, alongside model capability. 

Our contributions are:

\begin{itemize}
\item \textbf{A cross-paradigm, whole-repository migration benchmark.}
We introduce \methodname, comprising 63 PyTorch$\to$JAX migration tasks spanning 18 model families across training and serving, with native checkpoint conversion included as part of the migration contract (\S\ref{sec:benchmark}).

\item \textbf{An execution-grounded functional-equivalence evaluation.}
We evaluate migrated repositories by executing them in a separate \texttt{torch}-free, offline verifier against references recorded from the original PyTorch repository, combining an independence gate with continuous numeric and behavioral similarity (\S\ref{sec:bench-grading}).

\item \textbf{\harnessname, a reference-grounded migration harness.}
We introduce a diagnose--migrate--verify loop that turns the task-visible PyTorch repository into legitimate development-time feedback while maintaining a strict separation from hidden benchmark state, improving reward from $0.649$ to $0.723$ (\S\ref{sec:harness}; \S\ref{sec:experiments}).

\end{itemize}

\section{Related Work}
\label{sec:related}

\textbf{LLM-Based Code Conversion Benchmarks.}
MigrationBench \cite{liu2025migrationbench} and RepoTransBench \cite{wang2024repotransbench} evaluate repository-level migration and codebase translation; \S\ref{app:related} surveys related work. These benchmarks largely target API, language, version, or general codebase conversion through reference similarity \cite{lu2021codexglue,ren2020codebleu} or small-program execution. Beyond surface- or snippet-level evaluation, \methodname measures execution-grounded
functional equivalence after cross-paradigm migration of whole ML repositories across SFT, DPO, PPO, and paged-attention
serving against frozen PyTorch references, with a hard independence gate.

\textbf{Agentic Framework Migration.}
\citet{liang2026agentic} combine curated in-context translations, immutable PyTorch execution oracles, and iterative self-debugging, evaluating operators, neural modules, and ten human-assessed repositories with limited intervention. Building on this execution-grounded migration work, \methodname standardizes automated
evaluation of complete training and serving repositories through continuous numeric and
behavioral scores and a separate independence gate (Table~\ref{tab:benchmark-comparison}). \harnessname organizes source-execution feedback into diagnosis, migration, and verification while excluding hidden benchmark state.

\textbf{Reference-Grounded Feedback and Self-Repair.}
\harnessname builds on feedback-guided code repair. Intrinsic self-correction and self-verification remain unreliable \cite{huang2024selfcorrect,hong2024selfverification}; external signals support execution-trace self-debugging \cite{chen2024selfdebug}, tool-interactive critique \cite{gou2024critic}, and multi-agent review \cite{qian2024chatdev}. Claude Code \cite{anthropic2026claudecode} and Codex \cite{openai2025codex} implement similar diagnose-and-repair loops. \harnessname grounds cross-framework diagnosis, migration, and verification in task-visible
PyTorch behavior, preserving the evaluation boundary by excluding frozen references,
hidden tests, verifier code, and rewards.

\section{\methodname}
\label{sec:benchmark}

\begin{figure}[t]
  \centering
  \includegraphics[width=0.96\linewidth]{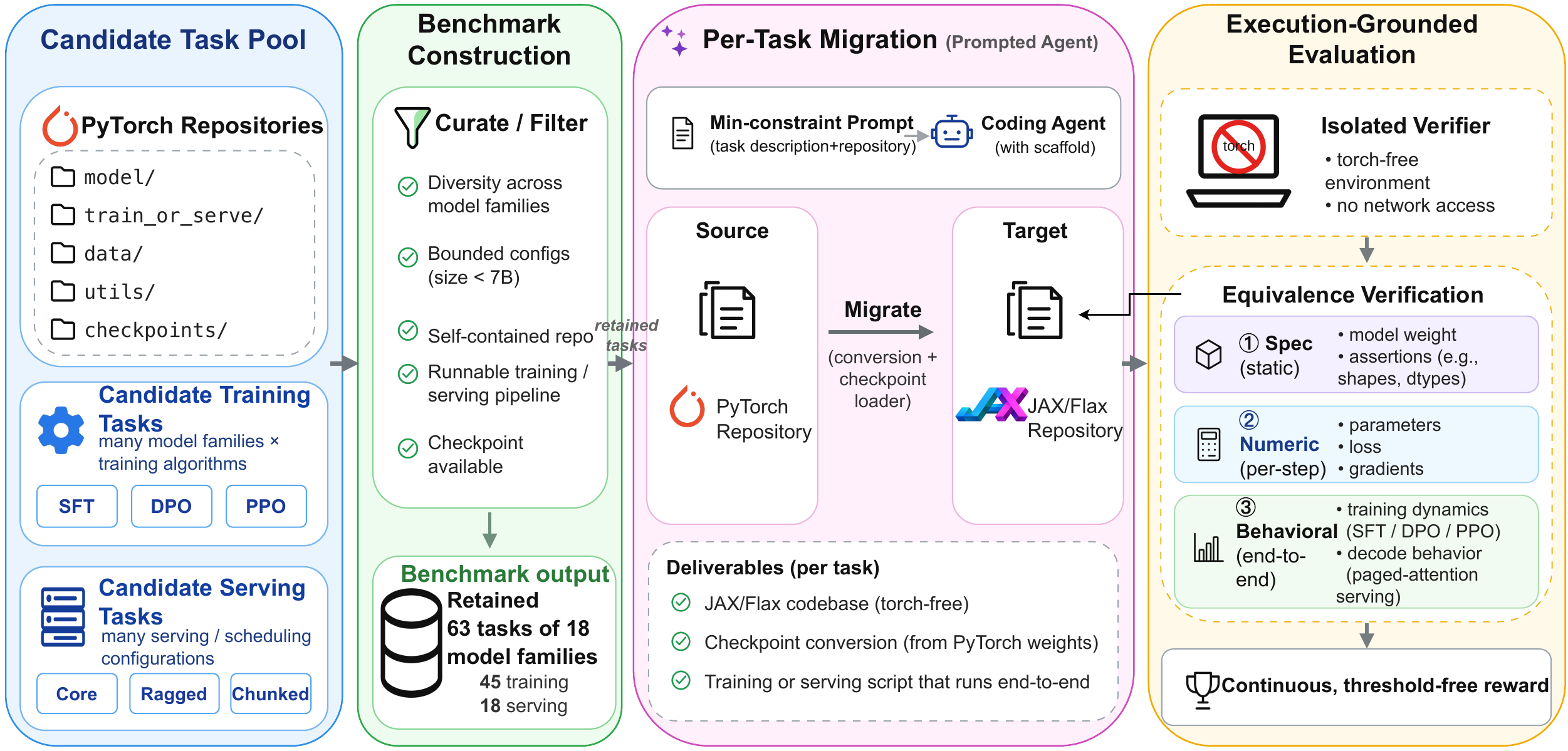}
\vspace{0pt}
\caption{\methodname overview: 63 PyTorch-to-JAX repository migrations across 18 model
families, covering 45 training and 18 paged-attention serving tasks. Agents deliver
native, \texttt{torch}-free JAX/Flax code, checkpoint conversion, and executable
training or serving paths. An offline verifier combines \textsc{Spec},
\textsc{Numeric}, and \textsc{Behavioral} checks into a continuous, threshold-free reward.}
  \label{fig:method-overview}
\vspace{0pt}
\end{figure}

\methodname comprises 63 whole-repository tasks that re-express eager, stateful PyTorch
modules as pure, compilable JAX/Flax functions
\citep{jax2018github,flax2020github,deepmind2020jax} for \emph{training} or \emph{inference
(serving)} migration. Candidates are graded by \emph{execution-grounded functional
equivalence} against recorded PyTorch references (Fig.~\ref{fig:method-overview}).

\subsection{The migration task}
\label{sec:bench-task}

Each task provides a self-contained, pure-PyTorch source tree $\mathcal{P}$ (roughly 15--20
modules from a real codebase) and the raw, publicly released checkpoint. The agent must
produce a native JAX reimplementation $\mathcal{J}$ that mirrors $\mathcal{P}$'s file layout
and public interface and writes its own PyTorch-free checkpoint-conversion routine ---
\emph{weight conversion is part of the deliverable}. The hard \emph{independence} constraint prohibits PyTorch use anywhere, even offline;
grading enforces it in a separate, offline, PyTorch-free environment.
The \emph{training} class is a $15\times3$ grid of 15 model families crossed with SFT, DPO
\citep{rafailov2023direct}, and PPO \citep{schulman2017proximal}, giving 45 LoRA
\citep{hu2021lora} fine-tuning tasks (24 text, 21 vision); the \emph{inference} class is a
$6\times3$ grid of six families crossed with three paged-attention engine modes of increasing
scheduling complexity (single-sequence, ragged-batch, chunked-prefill), giving 18 serving
tasks (15 text, 3 vision). Full family lists, source repositories, per-task requirements, and
modality details are in \S\ref{app:task-classes}.

\subsection{Grading design and machinery}
\label{sec:bench-grading}

\begin{table}[t]
\centering
\small
\caption{The five grading challenges, the principle that answers each, and the mechanism
that realizes it in the grader. The challenges and principles are stated in full in
\S\ref{app:design-challenges} and \S\ref{app:design-principles}.}
\vspace{0pt}
\label{tab:design-map}
\begin{tabular}{@{}p{0.85cm}p{2.5cm}p{2.5cm}p{7.0cm}@{}}
\toprule
 & Challenge & Principle & Mechanism in the grader \\
\midrule
C1\,/\,P1 & Fabricated equivalence & Independence-first & The grader runs offline with \texttt{torch} removed, so any hidden PyTorch import fails loudly; independence is a multiplicative gate that zeros the whole score \\
C2\,/\,P2 & ``Looks right'' $\neq$ ``behaves right'' & Execution-grounding & Every equivalence claim is settled by \emph{running} the candidate on fixed inputs --- weight conversion, forward, backward, and optimizer loop, or prefill and paged decode --- never by reading its source \\
C3\,/\,P3 & Uncertain ground truth & Oracle provenance & References are recorded once from the pure-PyTorch original; the reference JAX solution is graded as a candidate, never used as the ground truth \\
C4\,/\,P4 & No clean pass/fail & Bounded, threshold-free metrics & Each numeric and behavioral check returns a continuous $[0,1]$ similarity from an intrinsically bounded metric, with no tuned pass/fail cutoff \\
C5\,/\,P5 & Performance is part of correctness & Runtime as a graded dimension & The end-to-end levels (\textbf{Behavioral}) run under a uniform time budget; a JAX reimplementation that overruns it scores $0$ on those levels \\
\bottomrule
\end{tabular}
\vspace{0pt}
\end{table}

Correctness requires $\mathcal{J}$ to reproduce $\mathcal{P}$'s interface and behavior on the
same inputs: forward passes, gradients, optimizer trajectories, and trained-model
generalization for training; prefill/decode logits, sampling, and multi-request paged-cache
behavior for serving. This \emph{functional equivalence}, beyond test-pass counts, motivates
the five challenge--principle pairs in Table~\ref{tab:design-map}
(Appendix~\ref{app:design-challenges}--\ref{app:design-principles}).

The pipeline (\S\ref{app:lattice}--\S\ref{app:behavioral}) grades seven class-specific levels
(L0--L6, the \emph{lattice}): independence and structure (L0); module/operator and
composed-forward numerics across input shapes (L1--L3); loss and gradients (training) or step-wise
decoding (serving) at L4; and end-to-end execution and held-out behavior of trained artifacts or live
decode engines (L5--L6). Each level requires the preceding levels; scores reflect the level
reached rather than a single pass/fail verdict.

The levels feed three reward dimensions: L0 is \textbf{Spec}
(structure, interface, and independence; pass/fail), the single-pass levels L1--L4 feed
\textbf{Numeric}, and the whole-run levels L5--L6 feed \textbf{Behavioral} (L4's step-wise
checks also contribute to \textbf{Behavioral}). The dimensions combine into a spec-gated,
dimension-weighted reward:
\begin{equation}
R \;=\; \underbrace{s_{\text{spec}}}_{\text{multiplicative gate}}
\;\times\;
\bigl(w_{\text{num}}\, s_{\text{num}} + w_{\text{beh}}\, s_{\text{beh}}\bigr),
\qquad w_{\text{num}} = w_{\text{beh}} = 0.5 .
\label{eq:reward}
\end{equation}
where $R, s_{\text{spec}}, s_{\text{num}}, s_{\text{beh}}\in[0,1]$: $s_{\text{spec}}$ is the
fraction of \textbf{Spec} checks passed, and $s_{\text{num}}, s_{\text{beh}}$ are the mean
per-check similarity in the \textbf{Numeric} and \textbf{Behavioral} dimensions. The multiplicative Spec gate gives zero credit to reimplementations that are not genuine,
correctly shaped, and PyTorch-free, regardless of output similarity. Similarity measures
are intrinsically bounded and threshold-free, calibrated to the $\approx0.999$
floating-point noise ceiling. Any crash, PyTorch leak, NaN, or end-to-end run exceeding its
wall-clock budget (roughly twice the original's runtime) scores $0$ on the affected check
(\S\ref{app:lattice}, \S\ref{app:reward-details}, \S\ref{app:behavioral}).

\subsection{Benchmark validity and scope}
\label{sec:bench-quality}

We validate each task using its author-written, known-good \emph{reference JAX solution},
graded as a candidate, not ground truth. The full-pipeline \emph{oracle solvability check}
attains $0.980$ overall reward and $\approx0.999$ per-pass similarity, establishing task
solvability and reward calibration. A \emph{seeded-defect} sweep confirms failure on broken
code. High reward covers only graded properties, excluding fully-attended multimodal
numerics, exact token-by-token generation, and cross-hardware portability. Both safeguards
and the full non-goals list appear in \S\ref{app:validity}.

\section{\harnessname: A Reference-Grounded Harness}
\label{sec:harness}

\begin{figure}[t]
  \centering
  \includegraphics[width=0.96\linewidth]{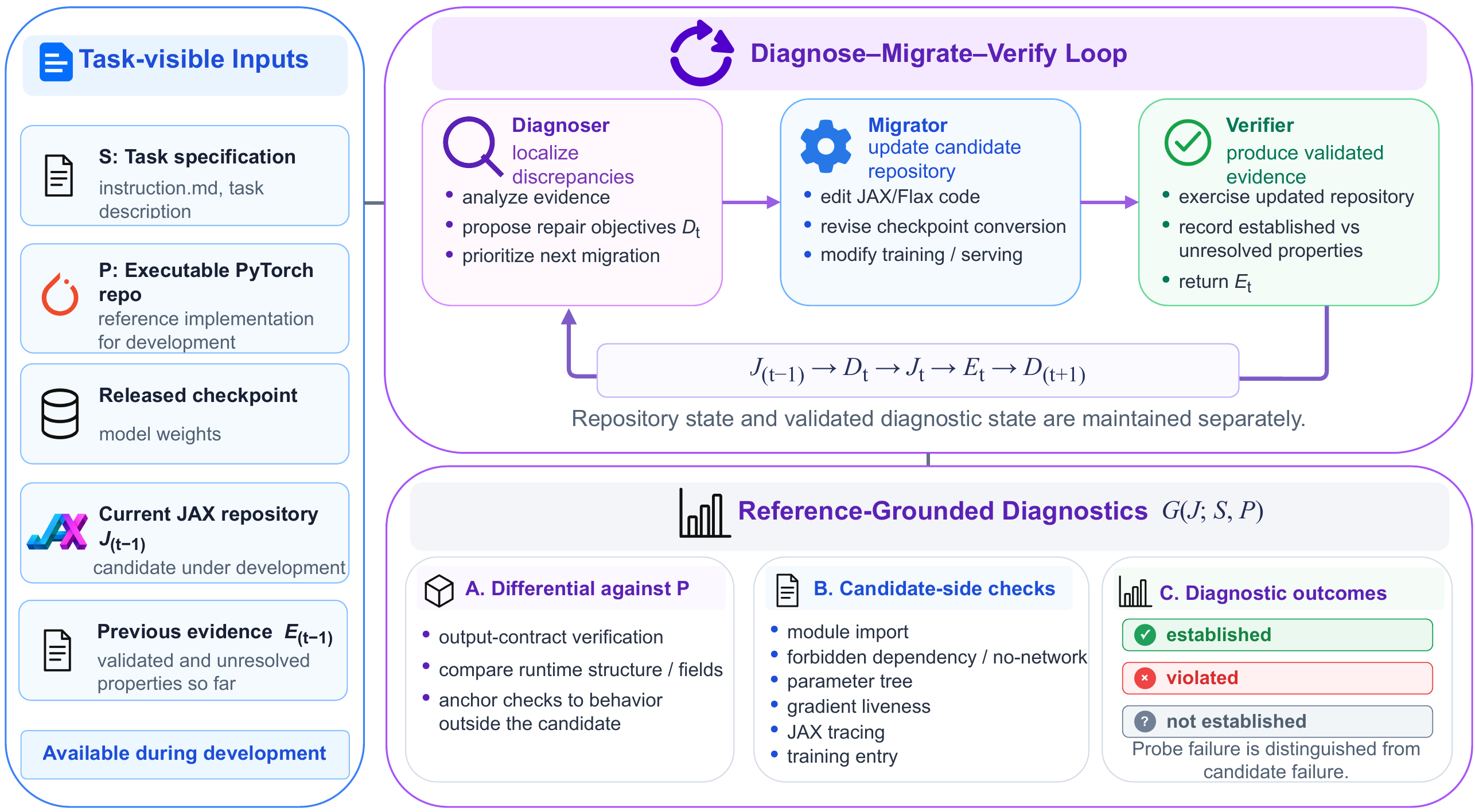}
\vspace{0pt}
\caption{\harnessname's diagnose--migrate--verify loop uses only task-visible artifacts.
Diagnostics $\mathcal{G}(\mathcal{J};\mathcal{S},\mathcal{P})$ combine differential
checks against the executable PyTorch source with candidate-side checks, labeling
evidence as \emph{established}, \emph{violated}, or \emph{not established}.}
  \label{fig:harness-overview}
\vspace{0pt}
\end{figure}

\harnessname grounds JAX migration in executable PyTorch behavior through a
\emph{diagnose--migrate--verify} loop (Figure~\ref{fig:harness-overview}). It combines
source comparisons with candidate-side checks and maintains repository state separately
from diagnostic evidence.

\subsection{Reference-Grounded Migration Loop}
\label{sec:harness-overview}

Let $\mathcal{S}$ be the task specification, $\mathcal{P}$ the executable PyTorch repository, and $\mathcal{J}_t$ the candidate JAX repository after iteration $t$. At each iteration, the \emph{Diagnoser} uses the current repository and prior evidence to identify remaining migration gaps $D_t$; the \emph{Migrator} updates the candidate; and the \emph{Verifier} executes the result to produce validated evidence $E_t$, following \mbox{$\mathcal{J}_{t-1}\to D_t\to\mathcal{J}_t\to E_t\to D_{t+1}$}. The loop separates implementation state $\mathcal{J}_t$ from evidence $E_t$ of established,
violated, or unresolved properties, avoiding diagnostic reconstruction solely from code
or interaction history. Unlike open-ended harness evolution, iteration reduces discrepancies between $\mathcal{P}$
and $\mathcal{J}_t$ against fixed, executable source behavior. Stage definitions appear in Appendix~\ref{app:harness-loop}.

\subsection{Reference-Grounded Self-Verification}
\label{sec:harness-grounding}

The development report $\mathcal{G}(\mathcal{J};\mathcal{S},\mathcal{P})$ records runtime
observations, structural evidence, and failure traces from task-visible execution, not
benchmark scores. Frozen PyTorch outputs belong exclusively to grading
(\S\ref{sec:bench-grading}); development executes $\mathcal{P}$ itself. Matched probes
compare $\mathcal{P}$ and $\mathcal{J}$'s output structure, exposed fields, and behavior,
anchoring verification outside the candidate to avoid circular self-validation.
Candidate-side checks cover executability, independence, parameter structure, gradient
liveness, JAX tracing, and integrated training or serving. Together these checks localize
divergence from repository validity through runtime interfaces to end-to-end behavior.
Outcomes are \emph{established}, \emph{violated}, or \textsc{not established}, separating
inconclusive probes from candidate errors (Appendix~\ref{app:harness-diagnostics}).

% \subsection{Information Boundary}
% \label{sec:harness-boundary}

% Reference-grounded verification is useful only if it preserves the benchmark's
% information boundary.
% \harnessname therefore uses only artifacts available to an ordinary task-solving agent:
% the task specification, the task-visible PyTorch repository, the released checkpoint,
% the evolving candidate repository, and outputs or exceptions produced by executing
% those artifacts.

% Conversely, \harnessname never reads hidden tests, frozen reference outputs, the verifier
% implementation, reference JAX solutions, benchmark rewards, scores, or per-check grading
% results.
% This separation also constrains diagnostic design: a diagnostic is included only when
% the property it examines follows from the task specification, the task-visible PyTorch
% repository, or the ordinary semantics of the exposed execution interface.
% Diagnostics whose design depends on hidden verifier behavior are excluded.
% The complete information boundary and its enforcement are detailed in
% Appendix~\ref{app:harness-boundary}.

% Finally, \harnessname changes only the procedure used to construct the candidate
% repository.
% Once migration terminates, the resulting repository is evaluated through the same
% unmodified \methodname grading pipeline as every off-the-shelf baseline.
% Neither the verifier nor the reward function is modified for \harnessname.
\section{Experiments}
\label{sec:experiments}

\subsection{Setup}
\label{sec:exp-setup}

\paragraph{Agents.}
We evaluate twenty combinations of five coding-agent scaffolds and their natively supported frontier models: \emph{claude-code} (Anthropic's CLI agent) with \texttt{claude-opus-5} and \texttt{claude-sonnet-5}; \emph{OpenHands}, a model-agnostic scaffold, with all six models; \emph{Gemini-CLI} (Google's CLI agent) with \texttt{gemini-3.1-pro}, \texttt{gemini-3.7-flash}, and \texttt{gemini-3.8-flash}; \emph{OpenCode}, an open-source CLI agent \citep{opencode}, with all six models; and \emph{Terminus-2}, Harbor's Terminus agent \citep{harbor}, with the three Gemini models. For compactness, Appendix~\ref{app:additional-baselines} (Table~\ref{tab:leaderboard-appendix})
reports OpenHands and Terminus-2; Table~\ref{tab:leaderboard} reports the remaining eleven
off-the-shelf configurations, \harnessname, and the oracle. Each agent receives only \texttt{instruction.md}, the pure-PyTorch source \texttt{torch\_orig/}, and the raw checkpoint (\S\ref{sec:benchmark}); it works in a networked container with one NVIDIA H100, 24 CPU cores, and 200\,GiB of RAM, under a 3-hour wall-clock budget per task, and emits \texttt{jax\_repo/}.

\paragraph{Grading and protocol.}
Every candidate is graded in the separate, no-network, \texttt{torch}-free verifier on a single pinned H100 (\S\ref{sec:benchmark}). We evaluate every experimental setting over the full 63-task suite with \emph{three random seeds}, including all baseline and \harnessname configurations at each reported refinement budget. Results are seed averages; reported standard deviations are sample standard deviations across seeds. The unmodified reference JAX solution serves as the \emph{oracle} candidate
(\S\ref{sec:bench-quality}), defining the faithful reimplementation ceiling.

\paragraph{Metrics.}
The primary metric is the continuous, spec-gated \textbf{reward} $R\in[0,1]$ of Eq.~\ref{eq:reward}, averaged over tasks. The reference JAX solution's aggregate reward of $0.980$ differs from the approximately
$0.999$ similarity in individual floating-point comparisons. Partial agreement earns partial credit subject to the spec gate. Three component scores localize failures: the \textbf{spec} pass rate (structural and independence correctness, the multiplicative gate), the \textbf{numeric} score (mean single-forward/backward similarity), and the \textbf{behavior} score (mean training-dynamics and invariant similarity). The leaderboard reports three-seed means; higher reward is better, while component scores
are diagnostics, not separately gated targets.

\subsection{Results}
\label{sec:exp-results}

\begin{table*}[t]
\centering
\small
\caption{Leaderboard on 63 tasks. $R$ is reward (Eq.~\ref{eq:reward});
\textsc{Spec}/\textsc{Num.}/\textsc{Beh.}\ are full-suite dimension scores;
\textsc{Train}/\textsc{Infer.}\ are rewards on 45 training/18 inference tasks.
All agent values, including \harnessname, are means over three random seeds;
$R$ subscripts give sample standard deviations.
The oracle is the reference JAX solution. Bold/underline mark the two highest
distinct means per column, excluding the oracle; ties share the same mark.
Additional baselines (OpenHands and Terminus-2) are reported in
Appendix~\ref{app:additional-baselines}.}
\vspace{-3pt}
\label{tab:leaderboard}
\begin{tabular}{llcccccc}
\toprule
Harness & Model & $R$ & \textsc{Spec} & \textsc{Num.} & \textsc{Beh.} & \textsc{Train} & \textsc{Infer.} \\
\midrule
Claude-Code & opus-5           & $0.649_{\,\pm.008}$ & 0.924 & 0.618 & 0.728 & 0.725 & 0.457 \\
Claude-Code & sonnet-5         & $0.408_{\,\pm.032}$          & 0.732 & 0.424 & 0.479 & 0.329 & 0.605 \\
\midrule
OpenCode    & opus-5           & $\underline{0.676}_{\,\pm.014}$ & \underline{0.927} & \underline{0.670} & \underline{0.741} & \underline{0.747} & 0.499 \\
OpenCode    & gemini-3.7-flash & $0.526_{\,\pm.033}$          & 0.883 & 0.592 & 0.535 & 0.595 & 0.351 \\
OpenCode    & sonnet-5         & $0.472_{\,\pm.042}$          & 0.760 & 0.467 & 0.554 & 0.390 & \textbf{0.677} \\
OpenCode    & glm-5.2          & $0.470_{\,\pm.051}$          & 0.732 & 0.495 & 0.544 & 0.491 & 0.418 \\
OpenCode    & gemini-3.8-flash & $0.361_{\,\pm.013}$          & 0.633 & 0.423 & 0.381 & 0.448 & 0.144 \\
OpenCode    & gemini-3.1-pro   & $0.064_{\,\pm.023}$          & 0.355 & 0.069 & 0.083 & 0.034 & 0.139 \\
\midrule
Gemini-CLI  & gemini-3.7-flash & $0.102_{\,\pm.020}$          & 0.346 & 0.120 & 0.111 & 0.081 & 0.155 \\
Gemini-CLI  & gemini-3.8-flash & $0.087_{\,\pm.012}$          & 0.324 & 0.100 & 0.102 & 0.069 & 0.131 \\
Gemini-CLI  & gemini-3.1-pro   & $0.057_{\,\pm.032}$          & 0.345 & 0.066 & 0.068 & 0.027 & 0.133 \\
\midrule
\rowcolor{gray!15}
\harnessname & opus-5          & $\textbf{0.723}_{\,\pm.022}$   & \textbf{0.931} & \textbf{0.730} & \textbf{0.792} & \textbf{0.764} & \underline{0.622} \\
\midrule
\multicolumn{2}{l}{\emph{Oracle}} & $0.980$ & 1.000 & 0.998 & 0.961 & 0.973 & 0.997 \\
\bottomrule
\end{tabular}
\vspace{-3pt}
\end{table*}

\begin{table*}[t]
\centering
\small
\caption{Reward by training method (15 tasks each) and inference mode (6 each).
Bold/underline mark the two highest distinct means per column, excluding the
oracle; ties share the same mark.}
\vspace{-3pt}
\label{tab:subtype}
\begin{tabular}{llcccccc}
\toprule
& & \multicolumn{3}{c}{Training method} & \multicolumn{3}{c}{Inference mode} \\
\cmidrule(lr){3-5}\cmidrule(lr){6-8}
Harness & Model & SFT & DPO & PPO & core & ragged & chunked \\
\midrule
Claude-Code & opus-5           & \textbf{0.854} & 0.784 & 0.538 & 0.489 & 0.472 & 0.409 \\
Claude-Code & sonnet-5         & 0.483 & 0.289 & 0.215 & \underline{0.626} & \textbf{0.604} & \underline{0.585} \\
\midrule
OpenCode    & opus-5           & 0.807 & \underline{0.839} & \underline{0.597} & \textbf{0.742} & 0.344 & 0.410 \\
OpenCode    & gemini-3.8-flash & 0.438 & 0.484 & 0.422 & 0.168 & 0.085 & 0.179 \\
\midrule
\rowcolor{gray!15}
\harnessname & opus-5          & \underline{0.821} & \textbf{0.851} & \textbf{0.620} & 0.623 & \underline{0.517} & \textbf{0.725} \\
\midrule
\multicolumn{2}{l}{\emph{Oracle}} & 0.966 & 0.965 & 0.988 & 0.999 & 0.996 & 0.995 \\
\bottomrule
\end{tabular}
\vspace{-3pt}
\end{table*}

% Table~\ref{tab:leaderboard} reports the main leaderboard; Table~\ref{tab:subtype} breaks reward down by training method and inference mode for the reported configurations and the oracle.

\paragraph{Higher reward, with substantial headroom.}
\harnessname leads five of six columns in Table~\ref{tab:leaderboard} in reported mean, including reward;
OpenCode with \texttt{sonnet-5} leads inference reward ($0.677$ vs.\ $0.622$). \harnessname exceeds the strongest off-the-shelf configuration, OpenCode with \texttt{opus-5}, by $0.047$ ($7.0\%$), from $0.676$ to $0.723$, and improves on Claude-Code with the same model by $0.074$ ($11.4\%$) across spec ($+0.007$), numeric ($+0.112$), behavior ($+0.064$), training ($+0.039$), and inference ($+0.165$). The gap to the $0.980$ oracle remains $0.257$.

\paragraph{Scaffold choice affects migration quality.}
With \texttt{opus-5}, OpenCode leads Claude-Code ($0.676$ vs.\ $0.649$), and with \texttt{gemini-3.8-flash} it leads Gemini-CLI ($0.361$ vs.\ $0.087$). Gemini-CLI's low spec scores ($0.324$--$0.346$) indicate frequent structural or independence failures. Section~\ref{sec:exp-reward-cost} examines the corresponding cost trade-offs.

\paragraph{Performance depends on both model and scaffold.}
\texttt{gemini-3.7-flash} outperforms \texttt{gemini-3.1-pro} under both Gemini-CLI and OpenCode, and model rankings depend on the task class: with Claude-Code, \texttt{opus-5} leads \texttt{sonnet-5} on training reward ($0.725$ vs.\ $0.329$), whereas \texttt{sonnet-5} leads on inference ($0.605$ vs.\ $0.457$). Performance thus depends on configuration; these comparisons do not separate reasoning
from tool-use effects.

\paragraph{Gains vary across task subtypes.}
Table~\ref{tab:subtype} shows \harnessname leading on DPO, PPO, and chunked inference, while Claude-Code retains the highest SFT (\texttt{opus-5}) and ragged (\texttt{sonnet-5}) scores and OpenCode with \texttt{opus-5} leads on core inference ($0.742$ vs.\ $0.623$). Relative to OpenCode with \texttt{opus-5}, \harnessname is higher on every subtype except core inference; relative to Claude-Code with \texttt{opus-5}, it improves five of six subtypes, most on chunked inference (from $0.409$ to $0.725$), but SFT decreases from $0.854$ to $0.821$. PPO remains the weakest training subtype for every configuration despite its $0.988$ oracle score, exposing a persistent gap in reproducing reinforcement-learning dynamics.

\paragraph{Takeaways.}
\harnessname improves aggregate training and inference reward, but its subtype gains are uneven and numeric and behavioral fidelity remain below the reference ceiling. We next examine these gains across costs, agent configurations, and refinement rounds.

\subsection{Reward--Cost Trade-offs under Continued Refinement}
\label{sec:exp-reward-cost}

\begin{figure}[t]
  \centering
  % Resolve the shared image height before subfigure changes \textwidth.
  \edef\comparisonPlotHeight{\the\dimexpr27\textwidth/100\relax}
  \begin{subfigure}[t]{0.45\textwidth}
    \centering
    \vspace{0pt}
    \includegraphics[height=\comparisonPlotHeight]{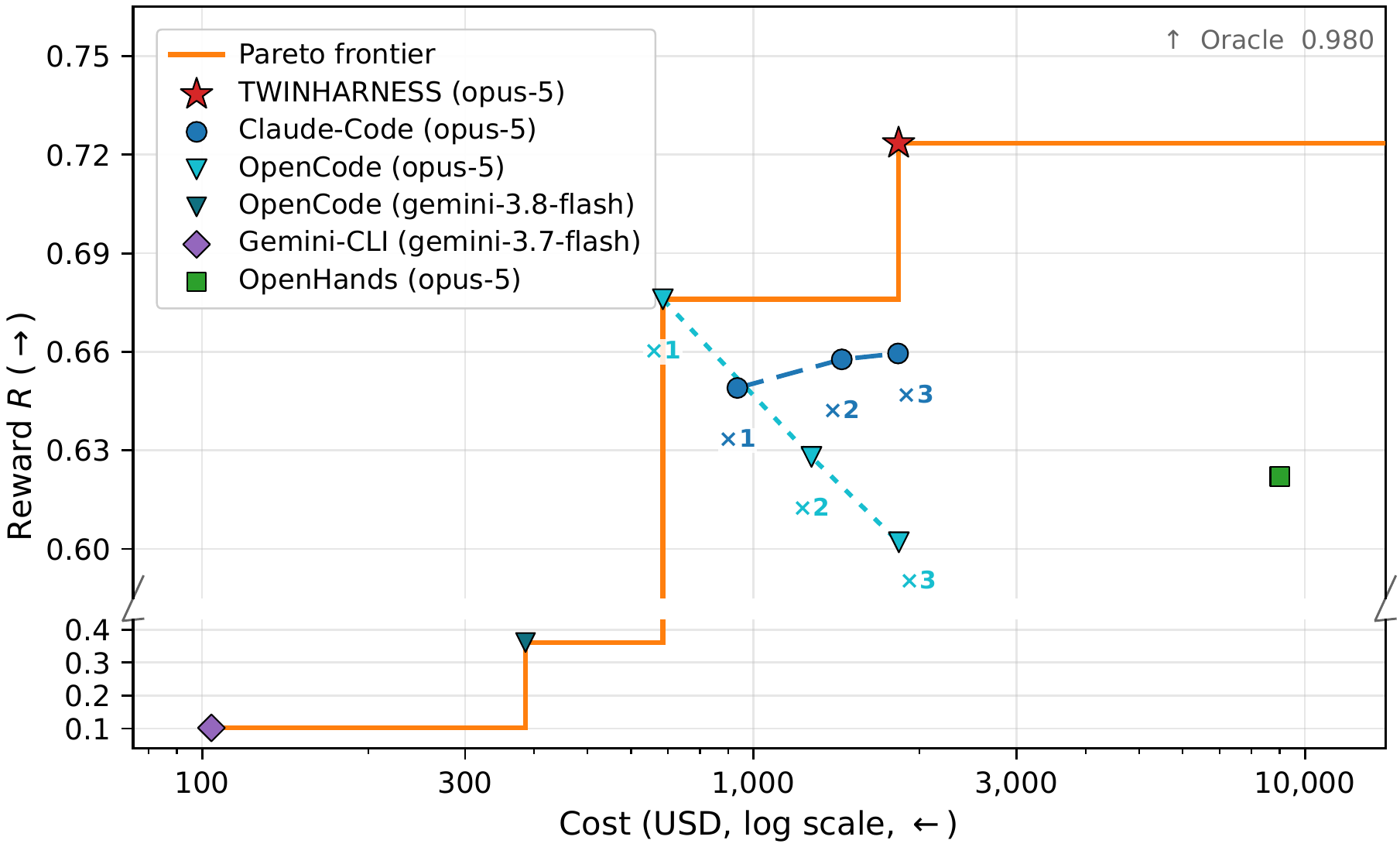}
    \caption{Reward--cost trade-offs.}
    \label{fig:reward-cost}
  \end{subfigure}\hfill
  \begin{subfigure}[t]{0.53\textwidth}
    \centering
    \vspace{0pt}
    \includegraphics[height=\comparisonPlotHeight]{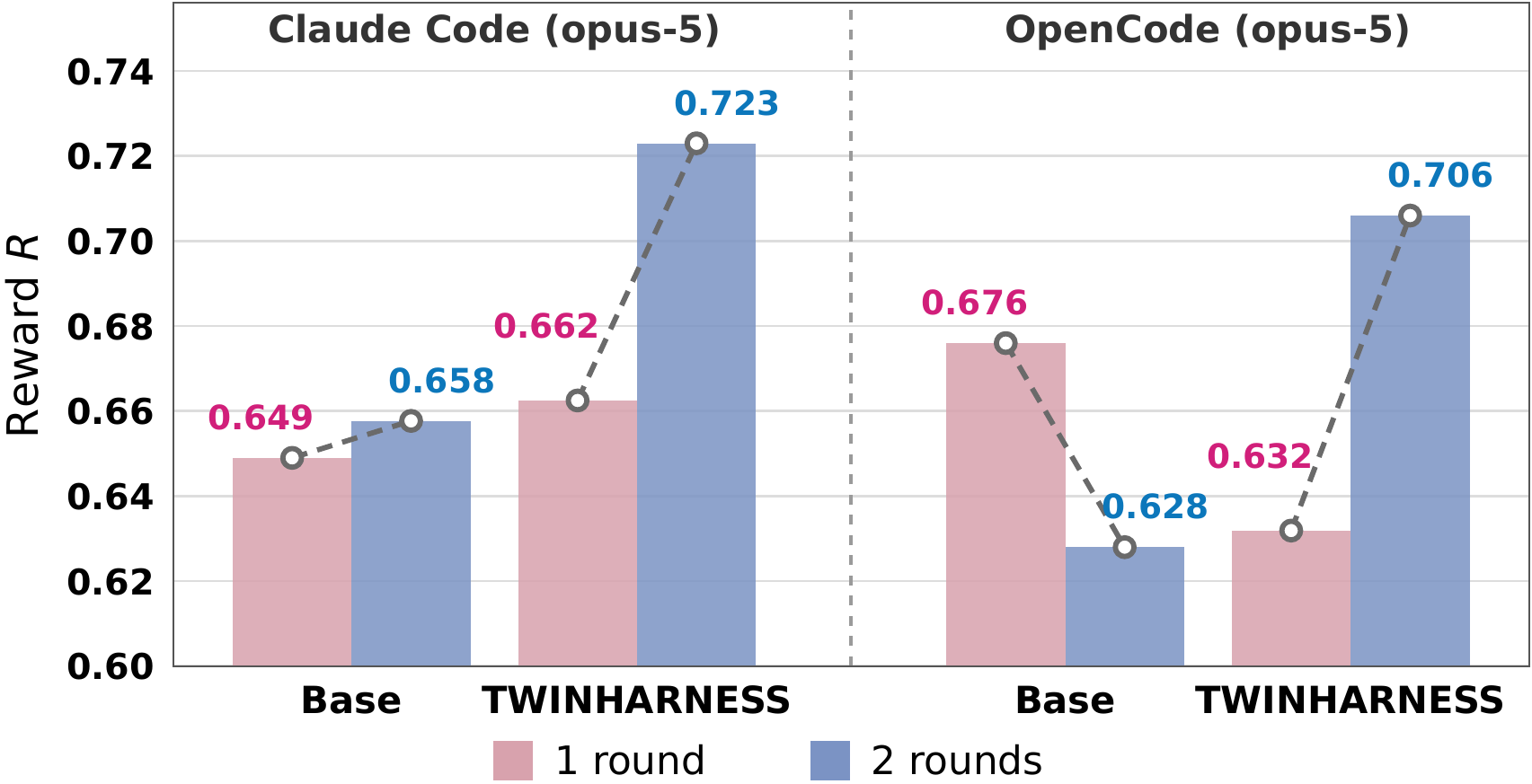}
    \caption{One versus two refinement rounds.}
    \label{fig:iteration-ablation}
  \end{subfigure}
\vspace{0pt}
\caption{\textbf{Cost and iteration trade-offs on \methodname.}
Rewards are means over three random seeds per setting.
\textbf{(a)} Reward versus API cost per 63-task run. Connected Claude-Code/OpenCode
points use \texttt{opus-5}; $\times1$ is the initial run, $\times2$/$\times3$ add
one/two continuations. Red star: \harnessname with \texttt{opus-5}; orange staircase:
best observed reward within each budget. Cost is logarithmic; the reward axis
is broken, with the $0.980$ oracle above range.
\textbf{(b)} One-round (pink) and two-round (blue) rewards with \texttt{opus-5}
on Claude-Code/OpenCode. Base uses ordinary continuation; \harnessname uses
diagnose--migrate--verify. Dashed lines connect each method's own trajectory.}
  \label{fig:cost-and-iteration}
\vspace{0pt}
\end{figure}

Figure~\ref{fig:reward-cost} compares reward and API cost per 63-task run, including
Claude-Code and OpenCode continuations with \texttt{opus-5}. Markers $\times1$, $\times2$,
and $\times3$ denote the initial run and one or two added continuations, respectively,
not the three random seeds used for statistical averaging or token-budget
multipliers. Baselines continue without
\harnessname's structured reference-grounded protocol.

\paragraph{A favorable trade-off beyond longer continuation.}
Initial OpenCode ($0.676$) has lower plotted API cost than \harnessname ($0.723$); both are non-dominated
in Figure~\ref{fig:reward-cost}. At an API cost close to that of \harnessname, two additional baseline continuations yield approximately $0.66$ for Claude-Code and $0.60$ for OpenCode, leaving reward gaps of about $0.06$ and $0.12$, respectively. Continuation does not close the gap, while initial OpenCode remains competitive at lower cost.

\paragraph{More refinement does not guarantee better migrations.}
Claude-Code improves modestly, whereas OpenCode falls from $0.676$ to $0.628$ after one continuation and to about $0.60$ after two, despite increasing API cost. Continued editing can thus degrade candidates, motivating structured verification feedback.
This does not isolate token-limit effects: continuation changes both edits and computation.

% AUTHOR CHECK: Confirm per-round budgets, continuation prompts,
% and cumulative API-cost accounting for the plotted refinement trajectories.

\begin{table*}[t]
\caption{Failure taxonomy on the same 63 tasks and verifier. Counts are three-seed means of task attempts
with at least one failed root-cause check per category; categories may overlap.
Downstream checks blocked by runtime errors are excluded. $\Delta$ is
\harnessname minus Claude Code; negative values indicate fewer failures.}
\vspace{0pt}
\label{tab:failure_taxonomy_rgh}
\centering
\scriptsize
\resizebox{\linewidth}{!}{%
\begin{tabular}{llccc}
\toprule
Category &
What goes wrong &
\harnessname &
Claude Code &
$\Delta$ \\
\midrule

& \multicolumn{4}{l}{\emph{Interface / admissibility (Spec stage)}} \\
\midrule

Init failure
& Candidate cannot be constructed under real configs
& 0 & 0.7 & $-0.7$ \\

Parameter-tree mismatch
& State-dict topology drifts from reference
& 30 & 28.7 & $+1.3$ \\

Batch-schema mismatch
& Collator output does not satisfy contract
& 10 & 7.7 & $+2.3$ \\

\midrule

& \multicolumn{4}{l}{\emph{Cross-paradigm runtime errors (Numeric stage)}} \\
\midrule

JAX/Torch boundary
& Type confusion at the JAX$\leftrightarrow$Torch seam
& 4 & 1.7 & $+2.3$ \\

Dtype unsupported
& Wrapper drops bf16 / mixed precision contract
& 6 & 5.3 & $+0.7$ \\

Artifact-contract drift
& Field exists but has wrong type/shape
& 0 & 1.0 & $-1.0$ \\

Missing artifact
& Required field absent from output
& 15 & 13.3 & $+1.7$ \\

Shape mismatch
& Token/label dims collapsed or expanded
& 1 & 7.3 & $-6.3$ \\

\midrule

& \multicolumn{4}{l}{\emph{Numeric disagreement (passes interface, disagrees on numbers)}} \\
\midrule

Forward mismatch
& \texttt{forward\_logits} or \texttt{forward\_loss} differs
& 4 & 2.7 & $+1.3$ \\

Method mismatch
& DPO/PPO/SFT method-specific tensors differ
& 15 & 17.0 & $-2.0$ \\

Gradient mismatch
& \texttt{gradient\_loss} / \texttt{gradient\_norm} differs
& 0 & 1.7 & $-1.7$ \\

\midrule

& \multicolumn{4}{l}{\emph{Behavioral disagreement / generation}} \\
\midrule

Generation mismatch / KV cache
& Sampled trajectory or KV state diverges
& 23 & 29.7 & $-6.7$ \\

\bottomrule
\end{tabular}%
}

\vspace{0pt}
\end{table*}

\subsection{Ablations of Agent Configuration and Iteration Budget}
\label{sec:exp-harness-ablations}

\paragraph{Applicability across scaffolds and model backbones.}
Across four matched scaffold--model pairs, \harnessname improves reward by $0.059$--$0.130$ (Appendix~\ref{app:harness-configurations}, Figure~\ref{fig:harness-backbones}). This supports applicability across the evaluated \methodname configurations, not transfer
to unseen task distributions.

\paragraph{Structured iteration versus ordinary continuation.}
Figure~\ref{fig:iteration-ablation} compares one- and two-round rewards for base agents and
\harnessname, fixing \texttt{opus-5} across Claude-Code and OpenCode. Base uses ordinary
continuation; \harnessname uses diagnose--migrate--verify. With Claude-Code, the base changes from $0.649$ to $0.658$ ($+0.009$), while \harnessname increases from $0.662$ to $0.723$ ($+0.061$). With OpenCode, the base decreases from $0.676$ to $0.628$ ($-0.048$), while \harnessname increases from $0.632$ to $0.706$ ($+0.074$). Thus, \harnessname trails the OpenCode base by $0.044$ after one round but leads by $0.078$ after two, also exceeding the initial base by $0.030$. The advantage emerges through iteration, not at every round budget. Larger one-to-two-round
gains are consistent with diagnostics aiding subsequent repairs, but neither establish
monotonic gains over further rounds nor isolate individual components. Section~\ref{sec:exp-reward-cost} complements this comparison with API costs.

\paragraph{Task-level refinement stability.}
With OpenCode, second-round regressions affect 2 of 63 tasks under \harnessname versus
22 under ordinary continuation (Appendix~\ref{app:refinement-stability}, Figure~\ref{fig:refinement-stability}). This is consistent with structured verification preserving achieved behavior, but does
not isolate component effects: each method starts from its own first-round candidate.

% AUTHOR CHECK: Confirm task coverage, round initialization,
% and stopping budgets for this ablation.

\subsection{A Cross-Agent Failure Taxonomy}
\label{sec:failure-taxonomy}

Under the same verifier and task set, \harnessname's improvements over Claude Code
concentrate on execution-visible failures (Table~\ref{tab:failure_taxonomy_rgh}): shape mismatches decrease by approximately $86\%$, generation/KV-cache failures by $23\%$, and no gradient mismatches are recorded for \harnessname. These runtime properties---tensor shapes, generation states, and gradient statistics---align
with the harness's additional reference-grounded feedback. Improvements are weaker for latent structural or semantic contracts: parameter-tree mismatches remain comparable, batch-schema and JAX/Torch-boundary failures increase, and method-specific numeric mismatches remain substantial. This suggests that \harnessname is most effective when correctness yields a locally executable
signal; hidden interface conventions and fine-grained semantic equivalence appear harder
to address with current feedback. The pattern is consistent with improved detection and repair of execution-visible incompatibilities,
not uniform error reductions, without isolating individual feedback components' causal contributions.

\section{Conclusion}
\label{sec:conclusion}

\methodname grades \emph{execution-grounded functional equivalence} in 63 whole-repository
PyTorch$\to$JAX migrations across 18 model families, covering training and serving.
A separate, \texttt{torch}-free verifier executes candidates against frozen PyTorch
references. The best of twenty off-the-shelf agent--model configurations scores $0.676$
against the $0.980$ oracle: high spec pass rates still leave substantial numeric and
behavioral gaps. Using task-visible PyTorch execution for legitimate self-verification,
\harnessname raises its \texttt{opus-5} scaffold's reward from $0.649$ to $0.723$, the
highest of any configuration (\S\ref{sec:harness}).
By evaluating whether migrated repositories preserve executable behavior, \methodname provides a foundation for measuring and advancing reliable cross-framework migration by coding agents.


\bibliographystyle{abbrvnat}
\nobibliography*
\bibliography{ref}

\begin{thebibliography}{54}
\providecommand{\natexlab}[1]{#1}
\providecommand{\url}[1]{\texttt{#1}}
\expandafter\ifx\csname urlstyle\endcsname\relax
  \providecommand{\doi}[1]{doi: #1}\else
  \providecommand{\doi}{doi: \begingroup \urlstyle{rm}\Url}\fi

\bibitem[Almeida et~al.(2024)Almeida, Xavier, and Valente]{almeida2024automatic}
A.~Almeida, L.~Xavier, and M.~T. Valente.
\newblock Automatic library migration using large language models: First results.
\newblock In \emph{Proceedings of the 18th ACM/IEEE International Symposium on Empirical Software Engineering and Measurement}, pages 427--433, 2024.
\newblock \doi{10.1145/3674805.3690746}.
\newblock URL \url{http://dx.doi.org/10.1145/3674805.3690746}.

\bibitem[Amin et~al.(2026)Amin, Fei, Li, Petke, and Ye]{amin2026jmigbench}
N.~Amin, Z.~Fei, X.~Li, J.~Petke, and H.~Ye.
\newblock {JMigBench}: A benchmark for evaluating {LLMs} on source code migration ({Java} 8 to {Java} 11).
\newblock In \emph{Proceedings of the 1st Workshop on Code Translation, Transformation, and Modernization (ReCode '26)}. ACM, 2026.
\newblock \doi{10.1145/3786180.3788316}.
\newblock URL \url{https://discovery.ucl.ac.uk/id/eprint/10222370/}.
\newblock In press.

\bibitem[{Anthropic}(2026)]{anthropic2026claudecode}
{Anthropic}.
\newblock {Claude Code} overview.
\newblock Anthropic Documentation, 2026.
\newblock URL \url{https://docs.anthropic.com/en/docs/claude-code/overview}.

\bibitem[Bradbury et~al.(2018)Bradbury, Frostig, Hawkins, Johnson, Katariya, Leary, Maclaurin, Necula, Paszke, Vander{P}las, Wanderman-{M}ilne, and Zhang]{jax2018github}
J.~Bradbury, R.~Frostig, P.~Hawkins, M.~J. Johnson, Y.~Katariya, C.~Leary, D.~Maclaurin, G.~Necula, A.~Paszke, J.~Vander{P}las, S.~Wanderman-{M}ilne, and Q.~Zhang.
\newblock {JAX}: composable transformations of {P}ython+{N}um{P}y programs, 2018.
\newblock URL \url{http://github.com/jax-ml/jax}.

\bibitem[Chan et~al.(2025)Chan, Chowdhury, Jaffe, Aung, Sherburn, Mays, Starace, Liu, Maksin, Patwardhan, et~al.]{chan2024mlebench}
J.~S. Chan, N.~Chowdhury, O.~Jaffe, J.~Aung, D.~Sherburn, E.~Mays, G.~Starace, K.~Liu, L.~Maksin, T.~Patwardhan, et~al.
\newblock Mle-bench: Evaluating machine learning agents on machine learning engineering.
\newblock In \emph{International Conference on Learning Representations}, volume 2025, pages 50466--50494, 2025.

\bibitem[Chen et~al.(2021)Chen, Tworek, Jun, Yuan, Pinto, Kaplan, Edwards, Burda, Joseph, Brockman, et~al.]{chen2021evaluating}
M.~Chen, J.~Tworek, H.~Jun, Q.~Yuan, H.~P. D.~O. Pinto, J.~Kaplan, H.~Edwards, Y.~Burda, N.~Joseph, G.~Brockman, et~al.
\newblock Evaluating large language models trained on code.
\newblock \emph{arXiv preprint arXiv:2107.03374}, 2021.

\bibitem[Chen et~al.(2024)Chen, Lin, Sch{\"a}rli, and Zhou]{chen2024selfdebug}
X.~Chen, M.~Lin, N.~Sch{\"a}rli, and D.~Zhou.
\newblock Teaching large language models to self-debug.
\newblock In \emph{The Twelfth International Conference on Learning Representations}, 2024.
\newblock URL \url{https://proceedings.iclr.cc/paper_files/paper/2024/hash/2460396f2d0d421885997dd1612ac56b-Abstract-Conference.html}.

\bibitem[Cheng et~al.(2025)Cheng, Shen, Yang, Wang, Cao, Ali, Wang, Hu, and Wang]{cheng2025codemenv}
K.~Cheng, X.~Shen, Y.~Yang, T.~Wang, Y.~Cao, M.~A. Ali, H.~Wang, L.~Hu, and D.~Wang.
\newblock {CODEMENV}: Benchmarking large language models on code migration.
\newblock In \emph{Findings of the Association for Computational Linguistics: ACL 2025}, pages 2719--2744. Association for Computational Linguistics, 2025.
\newblock \doi{10.18653/v1/2025.findings-acl.140}.
\newblock URL \url{https://aclanthology.org/2025.findings-acl.140/}.

\bibitem[DeepMind et~al.(2020)DeepMind, Babuschkin, Baumli, Bell, Bhupatiraju, Bruce, Buchlovsky, Budden, Cai, Clark, Danihelka, Dedieu, Fantacci, Godwin, Jones, Hemsley, Hennigan, Hessel, Hou, Kapturowski, Keck, Kemaev, King, Kunesch, Martens, Merzic, Mikulik, Norman, Papamakarios, Quan, Ring, Ruiz, Sanchez, Sartran, Schneider, Sezener, Spencer, Srinivasan, Stanojevi\'{c}, Stokowiec, Wang, Zhou, and Viola]{deepmind2020jax}
DeepMind, I.~Babuschkin, K.~Baumli, A.~Bell, S.~Bhupatiraju, J.~Bruce, P.~Buchlovsky, D.~Budden, T.~Cai, A.~Clark, I.~Danihelka, A.~Dedieu, C.~Fantacci, J.~Godwin, C.~Jones, R.~Hemsley, T.~Hennigan, M.~Hessel, S.~Hou, S.~Kapturowski, T.~Keck, I.~Kemaev, M.~King, M.~Kunesch, L.~Martens, H.~Merzic, V.~Mikulik, T.~Norman, G.~Papamakarios, J.~Quan, R.~Ring, F.~Ruiz, A.~Sanchez, L.~Sartran, R.~Schneider, E.~Sezener, S.~Spencer, S.~Srinivasan, M.~Stanojevi\'{c}, W.~Stokowiec, L.~Wang, G.~Zhou, and F.~Viola.
\newblock The {D}eep{M}ind {JAX} {E}cosystem, 2020.
\newblock URL \url{http://github.com/google-deepmind}.

\bibitem[Du et~al.(2025)Du, Liu, Guo, Wang, Huang, Ni, and Li]{du2025dependeval}
J.~Du, Y.~Liu, H.~Guo, J.~Wang, H.~Huang, Y.~Ni, and Z.~Li.
\newblock {DependEval}: Benchmarking {LLMs} for repository dependency understanding.
\newblock In \emph{Findings of the Association for Computational Linguistics: ACL 2025}, pages 7150--7179. Association for Computational Linguistics, 2025.
\newblock \doi{10.18653/v1/2025.findings-acl.373}.
\newblock URL \url{https://aclanthology.org/2025.findings-acl.373/}.

\bibitem[Gou et~al.(2024)Gou, Shao, Gong, Shen, Yang, Duan, and Chen]{gou2024critic}
Z.~Gou, Z.~Shao, Y.~Gong, Y.~Shen, Y.~Yang, N.~Duan, and W.~Chen.
\newblock {CRITIC}: Large language models can self-correct with tool-interactive critiquing.
\newblock In \emph{The Twelfth International Conference on Learning Representations}, 2024.
\newblock URL \url{https://openreview.net/forum?id=Sx038qxjek}.

\bibitem[{Harbor Framework Team}(2026)]{harbor}
{Harbor Framework Team}.
\newblock {Harbor: A framework for evaluating and optimizing agents and models in container environments}, 2026.
\newblock URL \url{https://doi.org/10.5281/zenodo.20953922}.

\bibitem[He et~al.(2025)He, Chen, Yang, Zhao, Yin, Kang, Lin, Rajmohan, and Zhang]{he2025execoder}
M.~He, Y.~Chen, F.~Yang, P.~Zhao, W.~Yin, Y.~Kang, Q.~Lin, S.~Rajmohan, and D.~Zhang.
\newblock {ExeCoder}: Empowering large language models with executability representation for code translation.
\newblock In \emph{Proceedings of the 2025 Conference on Empirical Methods in Natural Language Processing}, pages 7099--7125. Association for Computational Linguistics, 2025.
\newblock \doi{10.18653/v1/2025.emnlp-main.362}.
\newblock URL \url{https://aclanthology.org/2025.emnlp-main.362/}.

\bibitem[Heek et~al.(2026)Heek, Levskaya, Oliver, Ritter, Rondepierre, Steiner, and van {Z}ee]{flax2020github}
J.~Heek, A.~Levskaya, A.~Oliver, M.~Ritter, B.~Rondepierre, A.~Steiner, and M.~van {Z}ee.
\newblock {F}lax: A neural network library and ecosystem for {JAX}, 2026.
\newblock URL \url{http://github.com/google/flax}.

\bibitem[Hong et~al.(2024)Hong, Zhang, Pang, Yu, and Zhang]{hong2024selfverification}
R.~Hong, H.~Zhang, X.~Pang, D.~Yu, and C.~Zhang.
\newblock A closer look at the self-verification abilities of large language models in logical reasoning.
\newblock In \emph{Proceedings of the 2024 Conference of the North American Chapter of the Association for Computational Linguistics: Human Language Technologies (Volume 1: Long Papers)}, pages 900--925. Association for Computational Linguistics, 2024.
\newblock \doi{10.18653/v1/2024.naacl-long.52}.
\newblock URL \url{https://aclanthology.org/2024.naacl-long.52/}.

\bibitem[Hu et~al.(2021)Hu, Shen, Wallis, Allen-Zhu, Li, Wang, Wang, and Chen]{hu2021lora}
E.~J. Hu, Y.~Shen, P.~Wallis, Z.~Allen-Zhu, Y.~Li, S.~Wang, L.~Wang, and W.~Chen.
\newblock Lora: Low-rank adaptation of large language models.
\newblock \emph{arXiv preprint arXiv:2106.09685}, 2021.

\bibitem[Huang et~al.(2024)Huang, Chen, Mishra, Zheng, Yu, Song, and Zhou]{huang2024selfcorrect}
J.~Huang, X.~Chen, S.~Mishra, H.~S. Zheng, A.~W. Yu, X.~Song, and D.~Zhou.
\newblock Large language models cannot self-correct reasoning yet.
\newblock In \emph{The Twelfth International Conference on Learning Representations}, 2024.
\newblock URL \url{https://openreview.net/forum?id=IkmD3fKBPQ}.

\bibitem[Islam et~al.(2025)Islam, Jha, Mahmoud, Akhmetov, and Nadi]{islam2025empirical}
M.~Islam, A.~K. Jha, M.~Mahmoud, I.~Akhmetov, and S.~Nadi.
\newblock An empirical study of python library migration using large language models.
\newblock In \emph{2025 40th IEEE/ACM International Conference on Automated Software Engineering (ASE)}, pages 867--879. IEEE, 2025.
\newblock \doi{10.1109/ASE63991.2025.00077}.
\newblock URL \url{https://doi.org/10.1109/ASE63991.2025.00077}.

\bibitem[Islam et~al.(2024)Islam, Ali, and Parvez]{islam2024mapcoder}
M.~A. Islam, M.~E. Ali, and M.~R. Parvez.
\newblock {MapCoder}: Multi-agent code generation for competitive problem solving.
\newblock In \emph{Proceedings of the 62nd Annual Meeting of the Association for Computational Linguistics (Volume 1: Long Papers)}, pages 4912--4944, 2024.

\bibitem[Jimenez et~al.(2024{\natexlab{a}})Jimenez, Yang, Wettig, Yao, Pei, Press, and Narasimhan]{jimenez2023swebench}
C.~E. Jimenez, J.~Yang, A.~Wettig, S.~Yao, K.~Pei, O.~Press, and K.~Narasimhan.
\newblock {SWE}-bench: Can language models resolve real-world {GitHub} issues?
\newblock In \emph{The Twelfth International Conference on Learning Representations}, 2024{\natexlab{a}}.
\newblock URL \url{https://proceedings.iclr.cc/paper_files/paper/2024/hash/edac78c3e300629acfe6cbe9ca88fb84-Abstract-Conference.html}.

\bibitem[Jimenez et~al.(2024{\natexlab{b}})Jimenez, Yang, Wettig, Yao, Pei, Press, and Narasimhan]{jimenez2024swe}
C.~E. Jimenez, J.~Yang, A.~Wettig, S.~Yao, K.~Pei, O.~Press, and K.~Narasimhan.
\newblock Swe-bench: Can language models resolve real-world github issues?
\newblock In \emph{International Conference on Learning Representations}, volume 2024, pages 54107--54157, 2024{\natexlab{b}}.

\bibitem[Khan et~al.(2024)Khan, Bari, Do, Wang, Parvez, and Joty]{khan2024xcodeeval}
M.~A.~M. Khan, M.~S. Bari, X.~L. Do, W.~Wang, M.~R. Parvez, and S.~Joty.
\newblock {XCodeEval}: An execution-based large scale multilingual multitask benchmark for code understanding, generation, translation and retrieval.
\newblock In \emph{Proceedings of the 62nd Annual Meeting of the Association for Computational Linguistics (Volume 1: Long Papers)}, pages 6766--6805. Association for Computational Linguistics, 2024.
\newblock \doi{10.18653/v1/2024.acl-long.367}.
\newblock URL \url{https://aclanthology.org/2024.acl-long.367/}.

\bibitem[Kwon et~al.(2023)Kwon, Li, Zhuang, Sheng, Zheng, Yu, Gonzalez, Zhang, and Stoica]{kwon2023efficient}
W.~Kwon, Z.~Li, S.~Zhuang, Y.~Sheng, L.~Zheng, C.~H. Yu, J.~E. Gonzalez, H.~Zhang, and I.~Stoica.
\newblock Efficient memory management for large language model serving with pagedattention.
\newblock In \emph{Proceedings of the ACM SIGOPS 29th Symposium on Operating Systems Principles}, 2023.

\bibitem[Li et~al.(2025)Li, Zhang, Guo, Mao, Luo, Peng, Huang, Wang, and Li]{li2025feabench}
W.~Li, X.~Zhang, Z.~Guo, S.~Mao, W.~Luo, G.~Peng, Y.~Huang, H.~Wang, and S.~Li.
\newblock Fea-bench: A benchmark for evaluating repository-level code generation for feature implementation.
\newblock In \emph{Proceedings of the 63rd Annual Meeting of the Association for Computational Linguistics (Volume 1: Long Papers)}, pages 17160--17176. Association for Computational Linguistics, 2025.
\newblock \doi{10.18653/v1/2025.acl-long.839}.
\newblock URL \url{https://aclanthology.org/2025.acl-long.839/}.

\bibitem[Liang et~al.(2026)Liang, Ingram, Vanica, Gavrilescu, Harrat, Sipra, and Sankaran]{liang2026agentic}
Q.~Liang, S.~Ingram, G.~Vanica, A.~Gavrilescu, N.~Harrat, H.~Sipra, and S.~Sankaran.
\newblock Agentic framework for deep learning workload migration via in-context learning.
\newblock \emph{arXiv preprint arXiv:2606.15994}, 2026.

\bibitem[Liang et~al.(2025)Liang, Jiang, Hu, and Tan]{liang2025repocod}
S.~Liang, N.~Jiang, Y.~Hu, and L.~Tan.
\newblock Can language models replace programmers for coding? {REPOCOD} says {``}not yet{''}.
\newblock In \emph{Proceedings of the 63rd Annual Meeting of the Association for Computational Linguistics (Volume 1: Long Papers)}, pages 24698--24717. Association for Computational Linguistics, 2025.
\newblock \doi{10.18653/v1/2025.acl-long.1204}.
\newblock URL \url{https://aclanthology.org/2025.acl-long.1204/}.

\bibitem[Liu et~al.(2025{\natexlab{a}})Liu, Deng, Liu, Yang, Liu, Zhu, Zhao, Chai, Wu, JinKe, Zhang, Wang, Zhang, Tan, Xiang, Zhang, Su, and Zheng]{liu2025m2rceval}
J.~Liu, K.~Deng, C.~Liu, J.~Yang, S.~Liu, H.~Zhu, P.~Zhao, L.~Chai, Y.~Wu, J.~JinKe, G.~Zhang, Z.~M. Wang, G.~Zhang, Y.~Tan, B.~Xiang, Z.~Zhang, W.~Su, and B.~Zheng.
\newblock {M2RC-EVAL}: Massively multilingual repository-level code completion evaluation.
\newblock In \emph{Proceedings of the 63rd Annual Meeting of the Association for Computational Linguistics (Volume 1: Long Papers)}, pages 15661--15684. Association for Computational Linguistics, 2025{\natexlab{a}}.
\newblock \doi{10.18653/v1/2025.acl-long.763}.
\newblock URL \url{https://aclanthology.org/2025.acl-long.763/}.

\bibitem[Liu et~al.(2025{\natexlab{b}})Liu, Liu, Cheng, He, Shi, Guo, Zhu, Guo, Wang, and Wang]{liu2025repodebug}
J.~Liu, Z.~Liu, Z.~Cheng, M.~He, X.~Shi, Y.~Guo, X.~Zhu, Y.~Guo, Y.~Wang, and H.~Wang.
\newblock {RepoDebug}: Repository-level multi-task and multi-language debugging evaluation of large language models.
\newblock In \emph{Findings of the Association for Computational Linguistics: EMNLP 2025}, pages 23784--23813. Association for Computational Linguistics, 2025{\natexlab{b}}.
\newblock \doi{10.18653/v1/2025.findings-emnlp.1294}.
\newblock URL \url{https://aclanthology.org/2025.findings-emnlp.1294/}.

\bibitem[Liu et~al.(2025{\natexlab{c}})Liu, Pan, Xiang, He, Li, Du, and Gao]{liu2025projecteval}
K.~Liu, Y.~Pan, Y.~Xiang, D.~He, J.~Li, Y.~Du, and T.~Gao.
\newblock {ProjectEval}: A benchmark for programming agents automated evaluation on project-level code generation.
\newblock In \emph{Findings of the Association for Computational Linguistics: ACL 2025}, pages 20205--20221. Association for Computational Linguistics, 2025{\natexlab{c}}.
\newblock \doi{10.18653/v1/2025.findings-acl.1036}.
\newblock URL \url{https://aclanthology.org/2025.findings-acl.1036/}.

\bibitem[Liu et~al.(2025{\natexlab{d}})Liu, Liu, Zhou, Chen, Liu, Nguyen, Omidvar-Tehrani, Shen, Huan, Tripp, and Deoras]{liu2025migrationbench}
L.~Liu, X.~Liu, Q.~Zhou, L.~Chen, Y.~Liu, H.~Nguyen, B.~Omidvar-Tehrani, X.~Shen, J.~Huan, O.~Tripp, and A.~Deoras.
\newblock {MigrationBench}: Repository-level code migration benchmark from {Java} 8, 2025{\natexlab{d}}.
\newblock URL \url{https://arxiv.org/abs/2505.09569}.

\bibitem[Liu et~al.(2023)Liu, Xu, and McAuley]{liu2023repobench}
T.~Liu, C.~Xu, and J.~McAuley.
\newblock {RepoBench}: Benchmarking repository-level code auto-completion systems, 2023.
\newblock URL \url{https://arxiv.org/abs/2306.03091}.

\bibitem[Lu et~al.(2021)Lu, Guo, Ren, Huang, Svyatkovskiy, Blanco, Clement, Drain, Jiang, Tang, et~al.]{lu2021codexglue}
S.~Lu, D.~Guo, S.~Ren, J.~Huang, A.~Svyatkovskiy, A.~Blanco, C.~Clement, D.~Drain, D.~Jiang, D.~Tang, et~al.
\newblock Codexglue: A machine learning benchmark dataset for code understanding and generation.
\newblock \emph{arXiv preprint arXiv:2102.04664}, 2021.

\bibitem[Novikov et~al.(2025)Novikov, V{\~u}, Eisenberger, Dupont, Huang, Wagner, Shirobokov, Kozlovskii, Ruiz, Mehrabian, et~al.]{novikov2025alphaevolve}
A.~Novikov, N.~V{\~u}, M.~Eisenberger, E.~Dupont, P.-S. Huang, A.~Z. Wagner, S.~Shirobokov, B.~Kozlovskii, F.~J.~R. Ruiz, A.~Mehrabian, et~al.
\newblock {AlphaEvolve}: A coding agent for scientific and algorithmic discovery.
\newblock \emph{arXiv preprint arXiv:2506.13131}, 2025.

\bibitem[{OpenAI}(2025)]{openai2025codex}
{OpenAI}.
\newblock Introducing {Codex}.
\newblock OpenAI, 2025.
\newblock URL \url{https://openai.com/index/introducing-codex/}.

\bibitem[{OpenCode Team}(2025)]{opencode}
{OpenCode Team}.
\newblock {OpenCode: The open source AI coding agent}, 2025.
\newblock URL \url{https://opencode.ai/}.

\bibitem[Ou et~al.(2025)Ou, Liu, Chen, Wang, Peng, and Zheng]{ou2025rustrepotrans}
G.~Ou, M.~Liu, Y.~Chen, Y.~Wang, X.~Peng, and Z.~Zheng.
\newblock {RustRepoTrans}: Repository-level context code translation benchmark targeting {Rust}.
\newblock In \emph{40th IEEE/ACM International Conference on Automated Software Engineering (ASE 2025)}, pages 610--622. IEEE, 2025.
\newblock \doi{10.1109/ASE63991.2025.00057}.
\newblock URL \url{https://doi.org/10.1109/ASE63991.2025.00057}.

\bibitem[Paszke et~al.(2019)Paszke, Gross, Massa, Lerer, Bradbury, Chanan, Killeen, Lin, Gimelshein, Antiga, et~al.]{paszke2019pytorch}
A.~Paszke, S.~Gross, F.~Massa, A.~Lerer, J.~Bradbury, G.~Chanan, T.~Killeen, Z.~Lin, N.~Gimelshein, L.~Antiga, et~al.
\newblock Pytorch: An imperative style, high-performance deep learning library.
\newblock \emph{Advances in neural information processing systems}, 32, 2019.

\bibitem[Qian et~al.(2024)Qian, Liu, Liu, Chen, Dang, Li, Yang, Chen, Su, Cong, et~al.]{qian2024chatdev}
C.~Qian, W.~Liu, H.~Liu, N.~Chen, Y.~Dang, J.~Li, C.~Yang, W.~Chen, Y.~Su, X.~Cong, et~al.
\newblock {ChatDev}: Communicative agents for software development.
\newblock In \emph{Proceedings of the 62nd Annual Meeting of the Association for Computational Linguistics (Volume 1: Long Papers)}, pages 15174--15186, 2024.

\bibitem[Rafailov et~al.(2023)Rafailov, Sharma, Mitchell, Manning, Ermon, and Finn]{rafailov2023direct}
R.~Rafailov, A.~Sharma, E.~Mitchell, C.~D. Manning, S.~Ermon, and C.~Finn.
\newblock Direct preference optimization: Your language model is secretly a reward model.
\newblock \emph{Advances in neural information processing systems}, 36:\penalty0 53728--53741, 2023.

\bibitem[Ren et~al.(2020)Ren, Guo, Lu, Zhou, Liu, Tang, Sundaresan, Zhou, Blanco, and Ma]{ren2020codebleu}
S.~Ren, D.~Guo, S.~Lu, L.~Zhou, S.~Liu, D.~Tang, N.~Sundaresan, M.~Zhou, A.~Blanco, and S.~Ma.
\newblock Codebleu: a method for automatic evaluation of code synthesis.
\newblock \emph{arXiv preprint arXiv:2009.10297}, 2020.

\bibitem[Schulman et~al.(2017)Schulman, Wolski, Dhariwal, Radford, and Klimov]{schulman2017proximal}
J.~Schulman, F.~Wolski, P.~Dhariwal, A.~Radford, and O.~Klimov.
\newblock Proximal policy optimization algorithms.
\newblock \emph{arXiv preprint arXiv:1707.06347}, 2017.

\bibitem[Tang et~al.(2024)Tang, Kim, Song, Lothritz, Li, Ezzini, Tian, Klein, and Bissyand{\'e}]{tang2024codeagent}
X.~Tang, K.~Kim, Y.~Song, C.~Lothritz, B.~Li, S.~Ezzini, H.~Tian, J.~Klein, and T.~F. Bissyand{\'e}.
\newblock {CodeAgent}: Autonomous communicative agents for code review.
\newblock In \emph{Proceedings of the 2024 Conference on Empirical Methods in Natural Language Processing}, pages 11279--11313, 2024.

\bibitem[Trivedi et~al.(2024)Trivedi, Khot, Hartmann, Manku, Dong, Li, Gupta, Sabharwal, and Balasubramanian]{trivedi2024appworld}
H.~Trivedi, T.~Khot, M.~Hartmann, R.~Manku, V.~Dong, E.~Li, S.~Gupta, A.~Sabharwal, and N.~Balasubramanian.
\newblock {AppWorld}: A controllable world of apps and people for benchmarking interactive coding agents.
\newblock In \emph{Proceedings of the 62nd Annual Meeting of the Association for Computational Linguistics (Volume 1: Long Papers)}, pages 16022--16076, 2024.

\bibitem[Tschand et~al.(2026)Tschand, Hong, Walker, Cai, Wang, Subramanian, Dev, Reddi, Yazdanbakhsh, and Sankaran]{tschand2026jaxbench}
A.~Tschand, C.~Hong, J.~Walker, N.~Cai, S.~Wang, S.~Subramanian, S.~Dev, V.~J. Reddi, A.~Yazdanbakhsh, and S.~Sankaran.
\newblock Jaxbench: Benchmarking autonomous tpu kernel optimization.
\newblock \emph{arXiv preprint arXiv:2607.20466}, 2026.

\bibitem[{vLLM Project}(2026)]{tpuinference}
{vLLM Project}.
\newblock vllm tpu inference: A unified backend supporting pytorch and jax on tpu.
\newblock \url{https://github.com/vllm-project/tpu-inference}, 2026.

\bibitem[Wang et~al.(2025{\natexlab{a}})Wang, Liu, and Xiao]{wang2025cvebench}
P.~Wang, X.~Liu, and C.~Xiao.
\newblock {CVE-Bench}: Benchmarking {LLM}-based software engineering agents' ability to repair real-world {CVE} vulnerabilities.
\newblock In \emph{Proceedings of the 2025 Conference of the Nations of the Americas Chapter of the Association for Computational Linguistics: Human Language Technologies (Volume 1: Long Papers)}, pages 4207--4224. Association for Computational Linguistics, 2025{\natexlab{a}}.
\newblock \doi{10.18653/v1/2025.naacl-long.212}.
\newblock URL \url{https://aclanthology.org/2025.naacl-long.212/}.

\bibitem[Wang et~al.(2025{\natexlab{b}})Wang, Li, Song, Xu, Tang, Zhuge, Pan, Song, Li, Singh, Tran, Li, Ma, Zheng, Qian, Shao, Muennighoff, Zhang, Hui, Lin, Brennan, Peng, Ji, and Neubig]{wang2024openhands}
X.~Wang, B.~Li, Y.~Song, F.~F. Xu, X.~Tang, M.~Zhuge, J.~Pan, Y.~Song, B.~Li, J.~Singh, H.~H. Tran, F.~Li, R.~Ma, M.~Zheng, B.~Qian, D.~Shao, N.~Muennighoff, Y.~Zhang, B.~Hui, J.~Lin, R.~Brennan, H.~Peng, H.~Ji, and G.~Neubig.
\newblock {OpenHands}: An open platform for {AI} software developers as generalist agents.
\newblock In \emph{The Thirteenth International Conference on Learning Representations}, 2025{\natexlab{b}}.
\newblock URL \url{https://proceedings.iclr.cc/paper_files/paper/2025/hash/a4b6ad6b48850c0c331d1259fc66a69c-Abstract-Conference.html}.

\bibitem[Wang et~al.(2024)Wang, Wang, Wang, Guo, Chen, Grundy, Liu, Ma, Mao, Zhang, and Zheng]{wang2024repotransbench}
Y.~Wang, Y.~Wang, S.~Wang, D.~Guo, J.~Chen, J.~Grundy, X.~Liu, Y.~Ma, M.~Mao, H.~Zhang, and Z.~Zheng.
\newblock {RepoTransBench}: A real-world multilingual benchmark for repository-level code translation, 2024.
\newblock URL \url{https://arxiv.org/abs/2412.17744}.

\bibitem[Xia et~al.(2025)Xia, Deng, Dunn, and Zhang]{xia2024agentless}
C.~S. Xia, Y.~Deng, S.~Dunn, and L.~Zhang.
\newblock Demystifying {LLM}-based software engineering agents.
\newblock \emph{Proceedings of the ACM on Software Engineering}, 2\penalty0 (FSE):\penalty0 801--824, 2025.
\newblock \doi{10.1145/3715754}.
\newblock URL \url{https://doi.org/10.1145/3715754}.

\bibitem[Yan et~al.(2023)Yan, Tian, Li, Chen, and Wang]{yan2023codetransocean}
W.~Yan, Y.~Tian, Y.~Li, Q.~Chen, and W.~Wang.
\newblock {CodeTransOcean}: A comprehensive multilingual benchmark for code translation.
\newblock In \emph{Findings of the Association for Computational Linguistics: EMNLP 2023}, pages 5067--5089. Association for Computational Linguistics, 2023.
\newblock \doi{10.18653/v1/2023.findings-emnlp.337}.
\newblock URL \url{https://aclanthology.org/2023.findings-emnlp.337/}.

\bibitem[Yang et~al.(2024)Yang, Jimenez, Wettig, Lieret, Yao, Narasimhan, and Press]{yang2024sweagent}
J.~Yang, C.~E. Jimenez, A.~Wettig, K.~Lieret, S.~Yao, K.~Narasimhan, and O.~Press.
\newblock {SWE}-agent: Agent-computer interfaces enable automated software engineering.
\newblock In \emph{Advances in Neural Information Processing Systems 37}, 2024.
\newblock URL \url{http://papers.nips.cc/paper_files/paper/2024/hash/5a7c947568c1b1328ccc5230172e1e7c-Abstract-Conference.html}.

\bibitem[Zhang et~al.(2024)Zhang, Ruan, Fan, and Roychoudhury]{zhang2024autocoderover}
Y.~Zhang, H.~Ruan, Z.~Fan, and A.~Roychoudhury.
\newblock {AutoCodeRover}: Autonomous program improvement.
\newblock In \emph{Proceedings of the 33rd ACM SIGSOFT International Symposium on Software Testing and Analysis}, pages 1592--1604. ACM, 2024.
\newblock \doi{10.1145/3650212.3680384}.
\newblock URL \url{https://doi.org/10.1145/3650212.3680384}.

\bibitem[Zheng et~al.(2024)Zheng, Zhang, Zhang, Ye, and Luo]{zheng2024llamafactory}
Y.~Zheng, R.~Zhang, J.~Zhang, Y.~Ye, and Z.~Luo.
\newblock Llamafactory: Unified efficient fine-tuning of 100+ language models.
\newblock In \emph{Proceedings of the 62nd annual meeting of the association for computational linguistics (volume 3: system demonstrations)}, pages 400--410, 2024.

\bibitem[Ziftci et~al.(2025)Ziftci, Nikolov, Sj{\"o}vall, Kim, Codecasa, and Kim]{ziftci2025migrating}
C.~Ziftci, S.~Nikolov, A.~Sj{\"o}vall, B.~Kim, D.~Codecasa, and M.~Kim.
\newblock Migrating code at scale with {LLMs} at {Google}.
\newblock In \emph{Companion Proceedings of the 33rd ACM International Conference on the Foundations of Software Engineering}, pages 162--173. ACM, 2025.
\newblock \doi{10.1145/3696630.3728542}.
\newblock URL \url{https://doi.org/10.1145/3696630.3728542}.

\end{thebibliography}

%%%%%%%%%%%%%%%%%%%%%%%%%%%%%%%%%%%%%%%%%%%%%%%%%%%%%%%%%%%%%%%%%%%%%%%%%%%%%%%
%%%%%%%%%%%%%%%%%%%%%%%%%%%%%%%%%%%%%%%%%%%%%%%%%%%%%%%%%%%%%%%%%%%%%%%%%%%%%%%
% APPENDIX
%%%%%%%%%%%%%%%%%%%%%%%%%%%%%%%%%%%%%%%%%%%%%%%%%%%%%%%%%%%%%%%%%%%%%%%%%%%%%%%
%%%%%%%%%%%%%%%%%%%%%%%%%%%%%%%%%%%%%%%%%%%%%%%%%%%%%%%%%%%%%%%%%%%%%%%%%%%%%%%
\newpage
\appendix
\begin{center}
	\textbf{\LARGE Appendix }
\end{center}

% Appendix-only contents; entries and page numbers are generated automatically.
\startcontents[appendix]
\begingroup
\hypersetup{hidelinks}
\setcounter{tocdepth}{2}
\printcontents[appendix]{}{1}{\textbf{Contents of Appendix}\smallskip\hrule\smallskip}
\smallskip\hrule
\endgroup
\clearpage

\section{Additional Experimental Analyses}
\label{app:additional-experiments}

\subsection{Applicability Across Agent Configurations}
\label{app:harness-configurations}

\begin{figure}[htbp]
  \centering
  \includegraphics[width=0.70\linewidth]{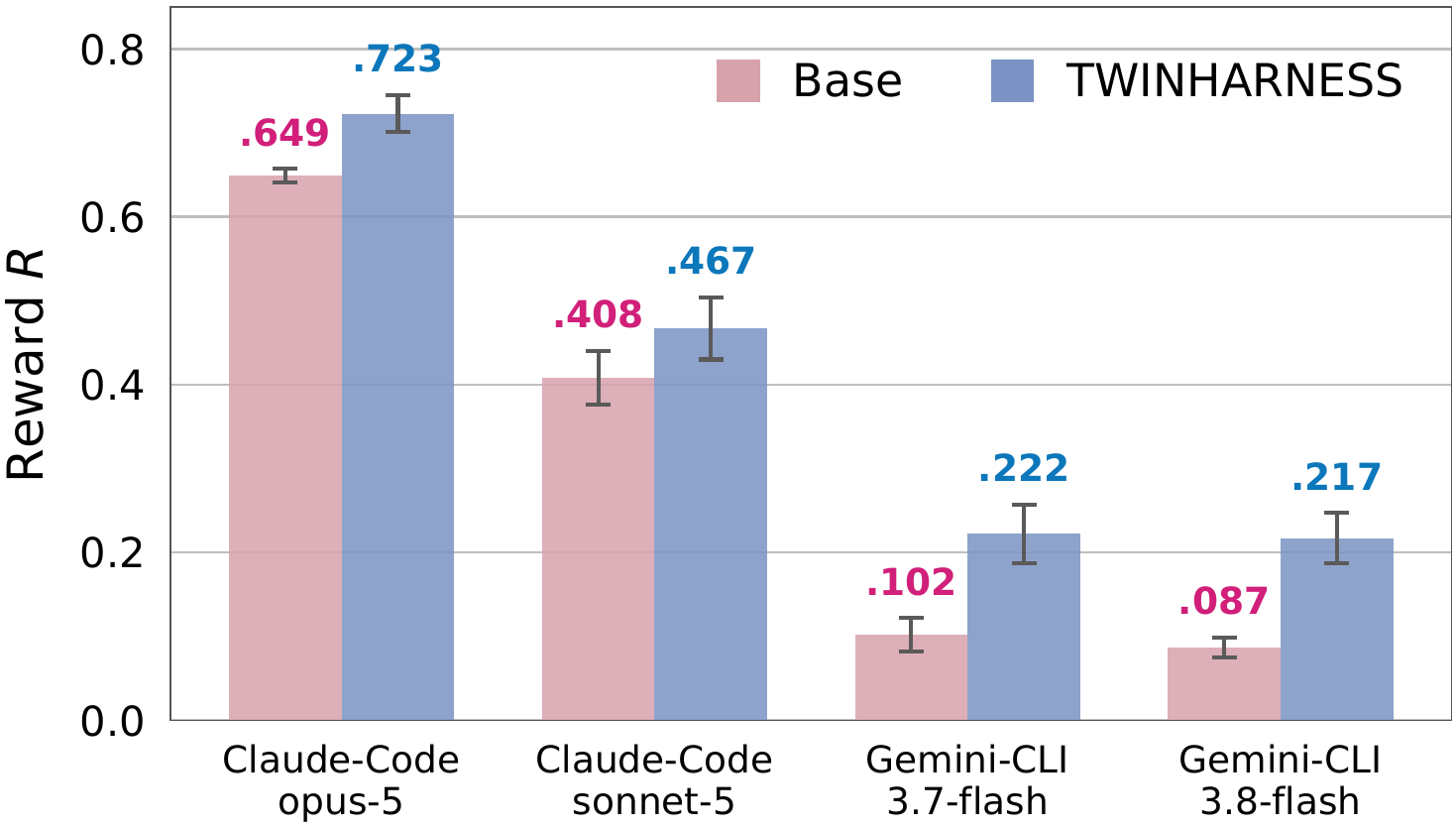}
\vspace{0pt}
\caption{\textbf{Reward across four scaffold--model pairs on \methodname.}
  Each pair compares the base agent (pink) with \harnessname (blue), holding
  the underlying scaffold and model fixed. Labels give mean rewards over three random seeds.}
  \label{fig:harness-backbones}
\vspace{0pt}
\end{figure}

Figure~\ref{fig:harness-backbones} compares the base agent with \harnessname for
four scaffold--model combinations. On Claude-Code, reward increases from
$0.649$ to $0.723$ with \texttt{opus-5} and from $0.408$ to $0.467$ with
\texttt{sonnet-5}, absolute gains of $0.074$ and $0.059$, respectively.
The Gemini-CLI configurations also show higher reported rewards, from $0.102$
to $0.222$ with \texttt{gemini-3.7-flash} and from $0.087$ to $0.217$ with
\texttt{gemini-3.8-flash}. Taken together, the
comparisons support applicability beyond a single scaffold--model pair on
\methodname, without establishing generalization to unseen task distributions.

% AUTHOR CHECK: Confirm the definition of the error bars.

\subsection{Task-Level Refinement Stability}
\label{app:refinement-stability}

\begin{figure}[htbp]
  \centering
  \includegraphics[width=0.78\linewidth]{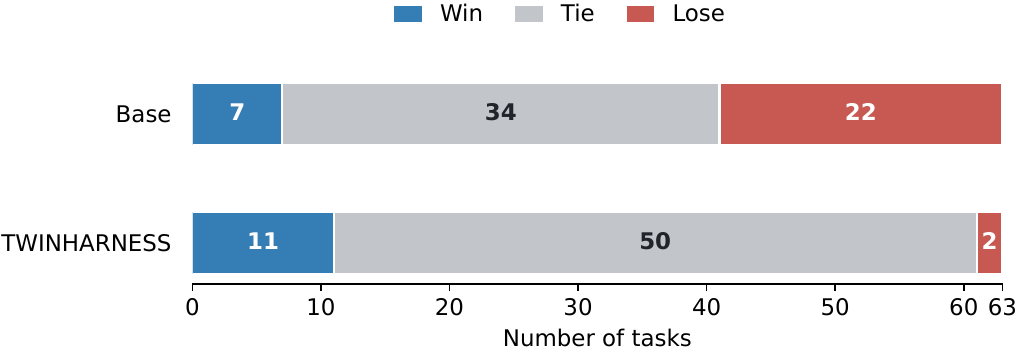}
\vspace{0pt}
\caption{\textbf{Task-level reward changes after a second round.}
  Results cover 63 tasks with OpenCode and \texttt{opus-5}, with
  $\Delta R=R^{(2)}-R^{(1)}$. Win: $\Delta R>0.01$;
  Tie: $|\Delta R|\leq0.01$; Lose: $\Delta R<-0.01$.
  Labels are task counts. Each method is compared with its own first-round candidate.}
  \label{fig:refinement-stability}
\vspace{0pt}
\end{figure}

Figure~\ref{fig:refinement-stability} examines task-level changes from the first
to the second round with OpenCode and \texttt{opus-5}, where
$\Delta R=R^{(2)}-R^{(1)}$. Ordinary continuation improves 7 of 63 tasks,
leaves 34 stable, and degrades 22 (34.9\%). In comparison, \harnessname
improves 11 tasks, leaves 50 stable, and degrades only 2 (3.2\%). Its
advantage therefore lies not only in improving more tasks, but also in
substantially reducing observed regressions during further refinement.

Execution-grounded feedback localizes discrepancies, while verification checks
repairs and informs subsequent diagnosis (Section~\ref{sec:harness-overview}).
The results are consistent with this feedback-guided refinement helping preserve
correct behavior. Because methods start from their own first-round artifacts,
these observations do not isolate individual components' causal effects.

% AUTHOR CHECK: Document the run/aggregation protocol for these 63-task counts.

\subsection{Additional Off-the-Shelf Baselines}
\label{app:additional-baselines}

To make the evaluation comprehensive, we provide additional off-the-shelf baseline results in this
section. Table~\ref{tab:leaderboard-appendix} reports two further scaffolds under the same protocol and
metrics as \S\ref{sec:exp-results}: OpenHands \citep{wang2024openhands} with all six models, and Terminus-2,
Harbor's Terminus agent \citep{harbor}, with the three Gemini models. All values are means over three random seeds on
the full 63-task suite, produced by the same unmodified grading pipeline (\S\ref{sec:bench-grading}). With
\texttt{opus-5}, OpenHands reaches $0.622$, below OpenCode ($0.676$) and Claude-Code ($0.649$) with the same
model (Table~\ref{tab:leaderboard}); Terminus-2 peaks at $0.369$ with \texttt{gemini-3.8-flash}. Every OpenHands and Terminus-2 configuration scores below both the top main-leaderboard baseline,
OpenCode with \texttt{opus-5} ($0.676$), and \harnessname ($0.723$). Including these baselines therefore
leaves the conclusions of \S\ref{sec:exp-results} unchanged: across all twenty off-the-shelf
configurations the highest reward is still $0.676$, and \harnessname remains the highest of any
configuration.

\begin{table}[htbp]
\centering
\small
\caption{Additional off-the-shelf baselines on 63 tasks, reported separately from the main leaderboard
(Table~\ref{tab:leaderboard}) under the same protocol and metrics. $R$ is reward (Eq.~\ref{eq:reward});
\textsc{Spec}/\textsc{Num.}/\textsc{Beh.}\ are full-suite dimension scores;
\textsc{Train}/\textsc{Infer.}\ are rewards on the 45 training/18 inference tasks. Values are means over three random
seeds; $R$ subscripts give sample standard deviations.}
\vspace{0pt}
\label{tab:leaderboard-appendix}
\begin{tabular}{llcccccc}
\toprule
Harness & Model & $R$ & \textsc{Spec} & \textsc{Num.} & \textsc{Beh.} & \textsc{Train} & \textsc{Infer.} \\
\midrule
OpenHands  & opus-5           & $0.622_{\,\pm.029}$ & 0.907 & 0.618 & 0.681 & 0.709 & 0.403 \\
OpenHands  & gemini-3.8-flash & $0.443_{\,\pm.016}$ & 0.828 & 0.555 & 0.435 & 0.525 & 0.235 \\
OpenHands  & gemini-3.7-flash & $0.416_{\,\pm.030}$ & 0.851 & 0.537 & 0.377 & 0.488 & 0.234 \\
OpenHands  & sonnet-5         & $0.401_{\,\pm.023}$ & 0.755 & 0.428 & 0.459 & 0.321 & 0.602 \\
OpenHands  & glm-5.2          & $0.247_{\,\pm.026}$ & 0.572 & 0.279 & 0.279 & 0.223 & 0.306 \\
OpenHands  & gemini-3.1-pro   & $0.086_{\,\pm.031}$ & 0.364 & 0.108 & 0.092 & 0.040 & 0.203 \\
\midrule
Terminus-2 & gemini-3.8-flash & $0.369_{\,\pm.015}$ & 0.793 & 0.496 & 0.353 & 0.453 & 0.160 \\
Terminus-2 & gemini-3.7-flash & $0.358_{\,\pm.012}$ & 0.786 & 0.479 & 0.331 & 0.442 & 0.147 \\
Terminus-2 & gemini-3.1-pro   & $0.006_{\,\pm.004}$ & 0.223 & 0.008 & 0.008 & 0.001 & 0.019 \\
\bottomrule
\end{tabular}
\vspace{0pt}
\end{table}

\clearpage
\section{Extended Related Work}
\label{app:related}

This appendix expands \S\ref{sec:related}, enumerating the benchmarks and systems the main
text cites only representatively.

\paragraph{LLM-based code conversion benchmarks.}
Conversion-oriented evaluation spans environment/version migration in CODEMENV
\cite{cheng2025codemenv}, Java migration in JMigBench \cite{amin2026jmigbench}, and
codebase-scale conversion in RustRepoTrans \cite{ou2025rustrepotrans}, alongside the
repository-level MigrationBench \cite{liu2025migrationbench} and RepoTransBench
\cite{wang2024repotransbench} cited in the main text. Adjacent work has examined library
conversion with LLMs through early empirical studies \cite{almeida2024automatic,islam2025empirical}
and large-scale industrial deployment settings \cite{ziftci2025migrating}. Related evaluation
has also grown through multilingual and execution-based benchmarks such as CodeTransOcean
\cite{yan2023codetransocean} and XCodeEval \cite{khan2024xcodeeval}, as well as conversion
methods evaluated at benchmark scale such as ExeCoder \cite{he2025execoder}. These efforts
substantially broaden evaluation beyond generic code generation, but they largely target API,
language, version, or general codebase conversion, and grade it either by surface-level
similarity to a reference \cite{lu2021codexglue,ren2020codebleu} or by executing small,
self-contained programs, rather than by execution-grounded equivalence of a whole running ML
system.

\subsection{Coding Agents and Codebase-Level Benchmarks}
\label{app:coding-agents}
A parallel line studies LLM-based agents for codebase-level software engineering,
established by issue resolution on SWE-bench \cite{jimenez2023swebench} and pursued by
scaffolds such as SWE-agent \cite{yang2024sweagent} and OpenHands \cite{wang2024openhands}. These evaluate completion, feature implementation, dependency
understanding, debugging, and ML-engineering workflows, but do not center on cross-framework
repository conversion under a fixed equivalence contract. \methodname complements them by
evaluating whether such agents can perform semantic-preserving, cross-paradigm conversion of
whole PyTorch repositories---across both training and serving---graded by execution-grounded
functional equivalence rather than issue resolution or code completion.

Additional scaffolds such as AutoCodeRover \cite{zhang2024autocoderover} and
Agentless \cite{xia2024agentless} explore different interfaces and pipelines for
codebase navigation, localization, editing, and execution.
Related agentic settings include multi-agent competitive code generation in MapCoder
\cite{islam2024mapcoder}, autonomous code review in CodeAgent \cite{tang2024codeagent},
interactive agent benchmarking in AppWorld \cite{trivedi2024appworld}, and broader scientific
and algorithmic discovery through coding agents in AlphaEvolve \cite{novikov2025alphaevolve}.
In parallel, a broader benchmark ecosystem has emerged around codebase-level coding ability,
including RepoBench \cite{liu2023repobench}, FEA-Bench \cite{li2025feabench}, RepoCod
\cite{liang2025repocod}, DependEval \cite{du2025dependeval}, ProjectEval \cite{liu2025projecteval},
RepoDebug \cite{liu2025repodebug}, CVE-Bench \cite{wang2025cvebench}, M2RC-EVAL
\cite{liu2025m2rceval}, and MLE-bench \cite{chan2024mlebench}. These evaluate codebase
completion, feature implementation, dependency understanding, debugging, vulnerability repair,
multilingual codebase reasoning, and ML engineering workflows, but do not center on
cross-framework repository conversion under a fixed equivalence contract.

\subsection{Autonomous Kernel Optimization}
\label{app:kernel-optimization}
JAXBench \citep{tschand2026jaxbench} evaluates autonomous Pallas kernel optimization on 50 JAX workloads for TPUs through numerical correctness and speedup over XLA. Its execution-feedback-driven operator optimization complements \methodname's repository-level PyTorch-to-JAX migration, which requires checkpoint conversion and faithful training or serving behavior.

\section{Benchmark design details}
\label{app:benchmark-design}

This appendix expands \S\ref{sec:benchmark}: the per-class task mechanics
(\S\ref{app:task-classes}), the full statements of the grading challenges
(\S\ref{app:design-challenges}) and design principles (\S\ref{app:design-principles}), the
equivalence lattice (\S\ref{app:lattice}), the continuous-fidelity reward
(\S\ref{app:reward-details}), the behavioral and runtime gates (\S\ref{app:behavioral}), and
the benchmark's calibration and documented boundaries (\S\ref{app:validity}).

\subsection{Task classes in detail}
\label{app:task-classes}

Each task gives the agent two things: (i)~a self-contained, pure-PyTorch source tree --- a
legible artifact of roughly 15--20 modules (about 4k lines) extracted from a real codebase
and stripped of heavy external dependencies --- and (ii)~the raw, publicly released model
checkpoint, provided inside the task environment. These source trees are not synthetic
snippets: the training-class sources are drawn from LLaMA-Factory
\citep{zheng2024llamafactory}, and the serving-class sources from vLLM
\citep{kwon2023efficient} and its JAX/TPU counterpart, tpu-inference \citep{tpuinference}, so
a passing JAX reimplementation is a migration of code that practitioners actually run. Each task is delivered
as a self-contained, containerized environment \citep{harbor} in two halves. The agent works
in a networked environment, with a GPU and both the original source and the checkpoint
available. The finished JAX reimplementation is then graded in a \emph{separate}, offline environment that
contains no PyTorch at all and receives only the agent's JAX repository together with the
same checkpoint. This split is the structural embodiment of the independence guarantee: the
agent may run PyTorch as a reference while working, but the graded JAX reimplementation runs where PyTorch
does not exist and the network is unavailable.

\paragraph{Training-migration class (45 tasks).}
The training class is a $15 \times 3$ grid: 15 model families crossed with three
post-training methods --- supervised fine-tuning (SFT), Direct Preference Optimization (DPO)
\citep{rafailov2023direct}, and Proximal Policy Optimization (PPO)
\citep{schulman2017proximal}. Every task is a LoRA \citep{hu2021lora} fine-tune (rank~8 on
all linear projections, base weights frozen; the trainable fraction is 0.28--0.88\%). The 15
families span eight text-only architectures (BLOOM, Falcon, Llama~3, Mistral, Phi-4, Qwen2,
Qwen3, StarCoder2) and seven multimodal ones (Gemma~3, InternVL, LLaVA-NeXT, PaliGemma,
Qwen2-VL, Qwen3-VL, Qwen2-Omni), giving 24 text-only and 21 vision-language tasks. The
benchmark grades only two modalities --- text and vision (image) --- and there is \emph{no
graded audio modality}: the single omni-modal family, Qwen2-Omni, is counted and graded as a
vision-language task, and its audio pathway is merely checked for presence rather than
numerically evaluated (\S\ref{app:validity}). To succeed, a JAX reimplementation must reproduce --- without
PyTorch --- the model's forward pass, the method-specific loss (shifted causal cross-entropy
for SFT, the preference objective for DPO, the policy surrogate for PPO), the AdamW optimizer
and its learning-rate schedule, the LoRA freeze/adapter structure, and the full data pipeline
(prompt templating, batching, and label masking). PPO tasks additionally run a reward model
inside the environment so that the whole loop stays self-contained. The three methods form a
natural difficulty gradient: SFT exercises a single supervised loss, DPO a paired preference
objective, and PPO a full sampling-and-update loop whose cross-framework reproducibility is
the loosest.

\paragraph{Inference-migration class (18 tasks).}
The inference class targets the serving end of the lifecycle: the agent reimplements a vLLM-style
paged-attention inference engine (about 1{,}500 lines over 19 modules) as a native JAX
implementation that preserves \emph{serving} semantics --- the paged key--value (KV) cache,
the block table that indexes it, the sampler's ordering of transforms, and greedy decoding
\citep{kwon2023efficient}. This is a distinct kind of migration object: state that PyTorch
mutates in place (the KV cache) must become an explicit value that the JAX code threads
through each step. The class is a $6 \times 3$ grid --- six families (Llama~3, Qwen2, Qwen3,
GPT-OSS, Gemma-4, and the vision-language Qwen2.5-VL) crossed with three engine modes of
increasing scheduling complexity: \emph{core} (single-sequence decode), \emph{ragged} (a
batched paged cache over requests of differing lengths), and \emph{chunked} (chunked prefill
with prefix reuse) --- giving 15 text and 3 vision-language tasks. Because the two classes
require different JAX versions (the serving stack needs a newer release), they build on
separate base environments but share the checkpoints and grading utilities.

\subsection{Design challenges}
\label{app:design-challenges}

Grading functional equivalence faithfully --- rather than rewarding a JAX reimplementation that merely
\emph{looks} like the original --- raises five challenges that ordinary code-translation
benchmarks do not confront. Table~\ref{tab:design-map} in the main text summarizes
their corresponding principles and grading mechanisms.

\paragraph{(C1) Fabricated equivalence.} A candidate can appear equivalent while secretly
relying on PyTorch, the original modeling libraries, or the network --- to load weights,
borrow a reference computation, or fetch something at run time --- so any numeric agreement
it reports may be unearned.

\paragraph{(C2) ``Looks right'' is not ``behaves right.''} A cross-paradigm rewrite turns
eager, stateful, object-oriented code into pure, compilable functions, and in-place cache
updates into explicitly threaded state; code that resembles the original, or that passes
hand-written unit tests, need not reproduce the system's numerics or dynamics.

\paragraph{(C3) Uncertain ground truth.} Grading a JAX reimplementation against a second, hand-written JAX
implementation would compare one implementation's quirks against another's; the target of
equivalence must instead be the original code that practitioners actually run.

\paragraph{(C4) No clean pass/fail.} Two independent 32-bit floating-point stacks disagree
slightly simply because they sum numbers in a different order, so exact equality is
impossible and any fixed tolerance is arbitrary and tunable; grading must nonetheless
separate a real bug from this floating-point noise and award partial credit across a large,
multi-stage system.

\paragraph{(C5) Performance is part of correctness.} A numerically perfect but too-slow JAX reimplementation
is not a usable migration: production training and paged-attention serving are viable only
within a compute budget, so a JAX reimplementation that reproduces every number yet cannot run at the
original's speed has not truly migrated the system.

\subsection{Design principles}
\label{app:design-principles}

We answer these challenges with five design principles that govern grading in \emph{both}
task classes (Table~\ref{tab:design-map}); the grading machinery each one drives is detailed
in \S\ref{app:lattice}--\S\ref{app:behavioral}.

\paragraph{(P1) Independence-first.} Numeric agreement is meaningless if the candidate
secretly calls PyTorch (C1), so independence is a precondition, not a checkbox. The grading
environment has no PyTorch, none of the original modeling libraries, and no network, and the
candidate is run in a setting where importing PyTorch simply fails. Any hidden dependency
therefore fails loudly rather than silently inflating a score --- whether it hides in a
training loop or in a decoding kernel.

\paragraph{(P2) Execution-grounding.} Because code that \emph{looks} like the original need
not \emph{behave} like it (C2), every equivalence claim is settled by \emph{running} the
candidate on fixed inputs, never by reading its source. A training candidate runs its own
weight conversion, forward pass, backward pass, and a real optimizer loop; a serving
candidate runs its own weight conversion, prefill, and multi-step paged decoding through its
own engine. In both cases the candidate is handed the reference's exact inputs and must
rebuild the model weights itself, so any observed agreement is attributable to the JAX reimplementation, not
to borrowed tensors or cached outputs.

\paragraph{(P3) Oracle provenance.} To fix an unambiguous ground truth (C3), the target of
equivalence is the \emph{pure-PyTorch original}, not a reference JAX solution. Reference
outputs are recorded from the PyTorch original on a fixed H100 GPU in true 32-bit precision
--- for training, the inputs, logits, gradients, loss trajectories, and held-out validation
losses; for serving, the prefill logits, per-step decode logits, greedy-decoded sequences,
and held-out-prompt log-probabilities --- and any change to the original forces them to be
re-recorded. The reference JAX solution is treated as a known-imperfect candidate, never as
the source of truth, which keeps the benchmark from grading one JAX reimplementation against another's
idiosyncrasies.

\paragraph{(P4) Bounded, threshold-free metrics.} Because no fixed tolerance can honestly
separate floating-point noise from a real bug (C4), every numeric and behavioral comparison
yields a \emph{continuous similarity in $[0,1]$} from an intrinsically bounded metric, rather
than a tuned pass/fail cutoff (\S\ref{app:reward-details}).

\paragraph{(P5) Runtime as a graded dimension.} Because a numerically perfect but too-slow
JAX reimplementation is not a usable migration (C5), speed is graded rather than assumed: the end-to-end
levels run under a uniform time budget derived from the PyTorch original, and a JAX reimplementation that
exceeds it scores $0$ on the affected check, exactly as a wrong one does
(\S\ref{app:behavioral}). Correctness in \methodname therefore includes running at the
original's speed, not merely returning its numbers.

\subsection{Equivalence lattice and aspects}
\label{app:lattice}

Equivalence is decomposed along a seven-level lattice (Table~\ref{tab:lattice}), read once
for each class: every level has a training reading and a serving reading, and a task earns
some prefix of the levels. Independently, each individual check is classified into exactly
one of three \emph{aspects} by the kind of evidence it provides: \textbf{Spec} (static
structure, public interface, parameter correspondence, and independence ---
hardware-invariant, pass/fail), \textbf{Numeric} (a single forward or backward pass, or an
isolated module or operator, on a fixed input, scored by continuous similarity), and
\textbf{Behavioral} (optimizer step loops, multi-step decode loops, reference-free
invariants, and the real end-to-end training or serving run). The three aspects are mutually
exclusive and cover every kind of evidence a check can give, and they are exactly the
dimensions the spec-gated reward (Eq.~\ref{eq:reward}) consumes. The lattice maps onto them
by depth: level~L0 is the multiplicative \textbf{Spec} gate, levels L1--L4 feed the
\textbf{Numeric} dimension (L4's step-wise mechanism checks also feed the \textbf{Behavioral}
one), and the whole-run levels L5--L6 feed \textbf{Behavioral}. In both classes, then, the
reward scores exactly the lattice prefix a task attains.

\begin{table*}[t]
\centering
\footnotesize
\caption{The seven-level functional-equivalence lattice, read per class. A task attains a
prefix; the top levels grade a \emph{trained model} (training) or a \emph{live paged-decode
engine} (serving), so a faithful JAX reimplementation must run end to end, not merely match a single forward
pass.}
\vspace{0pt}
\label{tab:lattice}
\begin{tabular}{@{}cp{1.9cm}p{4.15cm}p{4.15cm}c@{}}
\toprule
Level & Name & Training reading & Serving reading & Aspect \\
\midrule
L0 & Independence \& structure & PyTorch-free, compilable, mirrors the original's file and interface layout; converted parameters map one-to-one to the checkpoint. & PyTorch-free JAX engine mirroring the original's module and interface layout; converted parameters map one-to-one to the checkpoint. & Spec \\
L1 & Module / operator parity & every reimplemented submodule matches its PyTorch counterpart on the same input. & every reimplemented operator --- attention, rotary embeddings, the paged-attention kernel, sampler transforms --- matches its PyTorch counterpart. & Numeric \\
L2 & Composed-forward parity & the full forward pass (logits and hidden states) matches on the canonical batch. & the full prefill and a single decode step reproduce the reference logits on the canonical prompt. & Numeric \\
L3 & Generalization & forward parity holds across input shapes (length, ragged or padded, multiple rows). & logit parity holds across serving regimes: varying lengths, ragged batches, and chunked-prefill and prefix-reuse boundaries. & Numeric \\
L4 & Mechanism parity & the method's loss and its gradient match (value, authenticity, LoRA structure, learning-rate schedule). & greedy decoding reproduces the reference next-token distribution step by step (sampler ordering, decode/prefill consistency). & Num.\,/\,Beh. \\
L5 & End-to-end run parity & the JAX reimplementation's full training run reproduces the loss trend, optimizer identity, and decoding behavior. & the engine serves end to end with correct paging: invariance to batching, chunking, and prefix caching, determinism, and no leakage across sequences. & Behavioral \\
L6 & Held-out behavior parity & a model \emph{trained} by the JAX reimplementation reaches a held-out validation loss close to the PyTorch-trained model's. & greedy-decoded sequences and held-out-prompt log-probabilities match the reference engine's on \emph{unseen} prompts. & Behavioral \\
\bottomrule
\end{tabular}
\vspace{0pt}
\end{table*}

\subsection{Continuous-fidelity reward}
\label{app:reward-details}

Because two independent 32-bit floating-point stacks disagree only through differences in
summation order, a \emph{correct} JAX reimplementation does not score exactly $1.0$: single-pass comparisons
land at a measured noise ceiling of about $0.999$, and a real bug drives the score smoothly
below it, with no hand-set threshold separating ``exact'' from ``loose.'' Every comparison is
built from five intrinsically bounded similarity measures (Table~\ref{tab:primitives}), each
of which maps any not-a-number, infinity, or shape mismatch to $0$ so that a degenerate
output cannot look similar. Both operands are first promoted to 32-bit precision; the small
constant in each denominator ($\varepsilon = 10^{-12}$) only guards against division by zero
and cannot be tuned to change a verdict. The fixed factor of $2$ in the scalar measure is
likewise not a hidden tolerance: it rescales the symmetric difference to the true
mean-relative error, so the score reaches $0$ once two values differ by their own average
magnitude.

\begin{table}[t]
\centering
\small
\caption{The five similarity measures ($a\!\cdot\!b$ is a flattened dot product;
$\|\cdot\|$ is the $\ell_2$ norm; JSD is the Jensen--Shannon divergence). Every measure
returns a score in $[0,1]$ ($1$ = identical) and needs no threshold.}
\vspace{0pt}
\label{tab:primitives}
\begin{tabular}{lll}
\toprule
Measure & Definition & Applied to \\
\midrule
$S_{\text{scalar}}(a,b)$ & $1 - \dfrac{2\,|a-b|}{|a|+|b|+\varepsilon}$ & losses, log-probabilities, gradient norms, validation loss \\[1.2ex]
$S_{\ell_2}(a,b)$ & $1 - \dfrac{\|a-b\|}{\|a\|+\|b\|+\varepsilon}$ & magnitude-aware tensor closeness, schedules \\[1.2ex]
$S_{\cos}(a,b)$ & $\max\!\big(0,\ \tfrac{a\cdot b}{\|a\|\|b\|+\varepsilon}\big)$ & direction only (used inside $S_{\text{tensor}}$) \\[1.2ex]
$S_{\text{tensor}}(a,b)$ & $\sqrt{S_{\cos}\cdot S_{\ell_2}}$ & hidden states, module inputs/outputs, gradients \\[1.2ex]
$S_{\text{dist}}(p,q)$ & $1 - \dfrac{\mathrm{JSD}(\mathrm{softmax}\,p,\ \mathrm{softmax}\,q)}{\ln 2}$ & logits, any distribution over a vocabulary \\
\bottomrule
\end{tabular}
\vspace{0pt}
\end{table}

The measure is chosen by the \emph{shape} of the quantity being compared: distributions over
a vocabulary use the distributional measure (a Jensen--Shannon distance on the softmax, which
ignores the arbitrary per-token additive constant yet still reacts to temperature changes
that a scale-blind cosine would miss); single numbers use the scalar measure; and arrays use
the tensor measure, whose geometric mean of direction and magnitude ensures that a
uniform-scaling bug --- which a cosine alone would overlook --- still lowers the score.
Aggregation is two-level and threshold-free: within a multi-element check the score is the
plain average of the element scores, and each dimension's score is the plain average of its
checks' scores; nothing is rounded to pass/fail on the way up.

These per-check scores combine into the primary reward (Eq.~\ref{eq:reward}), a
\emph{spec-gated, dimension-weighted similarity}. Spec acts as a multiplicative
anti-fabrication gate: a JAX reimplementation that is not genuine, correctly shaped, and PyTorch-free
earns nothing, however close its numbers happen to look, and a partially
correct structure scales the entire reward by the fraction of spec checks it passes. The two
similarity dimensions each contribute the average of their per-check $[0,1]$ scores,
renormalized over whichever dimensions actually ran, so partial credit is continuous rather
than a series of stage pass/fails. Every failure mode maps to $0$ rather than being skipped:
a crash, a PyTorch leak, a not-a-number result, or an end-to-end run that exceeds the uniform
time budget (\S\ref{app:behavioral}) all score $0$ on the affected check --- too slow counts
for exactly as little as wrong.

\subsection{Behavioral gates and the runtime dimension}
\label{app:behavioral}

Realizing P5, the highest levels grade a \emph{trained artifact}, not a single forward pass.
Level~L6 splits each task's bundled dataset into fixed training and validation subsets (using
a fixed, framework- and platform-independent random shuffle of the indices), trains a model
using the candidate's own training entry point --- inside the PyTorch-blocked, offline
environment --- and scores the resulting held-out validation loss against the recorded
PyTorch reference using a \emph{banded} version of the scalar measure. The width of the band
is set by the \emph{reference's own} run-to-run variability (measured by re-running the
PyTorch original), never by the PyTorch-versus-JAX gap, so a faithful JAX reimplementation lands inside the
band and scores $1.0$, and only divergence beyond the reference's own fluctuation pulls the
score down. Because the small single-pass probes barely move the model, L6 shares one real
training run with the end-to-end level L5, and that run is capped by a uniform time budget
derived from the PyTorch reference's wall-clock time. A numerically perfect but too-slow JAX reimplementation
therefore scores $0$: clearing L5 and L6 requires the JAX reimplementation to be genuinely optimized for
speed --- compiled fixed-shape execution, fused kernels, and static-shape rewrites of eager
sub-networks --- which is what makes \emph{runtime a first-class graded dimension} rather
than an afterthought.

\subsection{Calibration and documented boundaries}
\label{app:validity}

Two safeguards protect the benchmark's own validity. First, a \emph{mirror check} statically
verifies that the reference JAX solution and the PyTorch original expose the same set of
modules and that every generated task instruction matches its template, so the specification
the agent reads never drifts from the reference. Second, an \emph{oracle solvability check}
submits the reference JAX solution itself as a candidate and grades it through the full
pipeline. This oracle attains a reward of $0.973$ on the training class (34 of 45 tasks pass
every check strictly) and $0.997$ on the inference class (18 of 18), for $0.980$ overall (52
of 63) --- with per-pass numeric similarity near $0.999$.
The overall reward aggregates structural, numeric, and behavioral checks and
should not be identified with this per-pass similarity. The reference results
provide an empirical baseline for faithful reimplementation, while the
remaining strict-check failures delimit the claim of exact agreement.
For the inference class, a
\emph{seeded-defect} sweep further shows that $84\%$ of the collected checks are tripped by
at least one deliberately injected bug in the reference, evidence that the checks have teeth
rather than merely staying green on correct code.

We state non-goals explicitly rather than letting a high reward quietly absorb them.
Multimodal grading pins the text-token forward pass and, by design, the vision tower and its
merge step up to the point where they enter the causal mask; but \emph{fully attended}
multimodal numerics (image content flowing all the way into a graded language-model logit)
and language-side submodule parity on multimodal families remain a scoped frontier, and audio
(for the omni-modal family) is checked only for presence rather than graded. Several other
items are deliberately out of scope: cache- and end-of-sequence-aware decoding within the
training class; exact token-by-token generation (tie-breaking among equal-probability tokens
and reduction-order flips make the exact token ids irreproducible, so the tie-robust
next-token distribution is scored instead); LoRA gradient \emph{magnitudes} beyond the first
step (which depend on random initialization once it differs across frameworks, so they are
scored structurally and through initialization-stable statistics); and portability of grading
across arbitrary hardware (the suite fixes a single H100 GPU so the choice of device never
enters a verdict). Naming these keeps the benchmark's coverage legible: what a high reward
proves is exactly the graded lattice prefix, and nothing is smuggled in as unearned coverage.

\section{Additional Details of \harnessname}
\label{app:harness-details}

This appendix provides the complete stage-wise formalization of the
\emph{diagnose--migrate--verify} loop, the full taxonomy of reference-grounded
diagnostics, and the information boundary enforced by \harnessname.

\subsection{Detailed Diagnose--Migrate--Verify Loop}
\label{app:harness-loop}

Let $\mathcal{S}$ denote the task specification, $\mathcal{P}$ the task-visible PyTorch
repository, and $\mathcal{J}_t$ the candidate JAX repository after iteration $t$.
We denote by
$\mathcal{G}(\mathcal{J};\mathcal{S},\mathcal{P})$
the reference-grounded diagnostic report introduced in
\S\ref{sec:harness-grounding}.

The \emph{Diagnoser} first analyzes the current candidate together with previously
validated evidence and identifies the behavioral discrepancies that should be addressed
next:
\begin{equation}
D_t =
\mathrm{Diagnose}\!\left(
    \mathcal{S},
    E_{t-1},
    \mathcal{G}(\mathcal{J}_{t-1};\mathcal{S},\mathcal{P});
    \mathrm{ro}(\mathcal{J}_{t-1})
\right).
\label{eq:harness-diagnose}
\end{equation}

The \emph{Migrator} then updates the JAX repository to resolve the diagnosed
cross-framework discrepancies:
\begin{equation}
\mathcal{J}_t =
\mathrm{Migrate}\!\left(
    \mathcal{J}_{t-1};
    \mathcal{S},
    D_t,
    \mathcal{G}(\mathcal{J}_{t-1};\mathcal{S},\mathcal{P})
\right).
\label{eq:harness-migrate}
\end{equation}

Finally, the \emph{Verifier} independently exercises the updated repository and records
which behaviors have been established and which remain unresolved:
\begin{equation}
E_t =
\mathrm{Verify}\!\left(
    \mathrm{ro}(\mathcal{J}_t);
    \mathcal{S},
    D_t,
    \mathcal{G}(\mathcal{J}_t;\mathcal{S},\mathcal{P})
\right).
\label{eq:harness-verify}
\end{equation}

Equations~(\ref{eq:harness-diagnose})--(\ref{eq:harness-verify}) formalize the three
stages of the migration loop.
Here $D_t$ denotes a diagnosis of the remaining migration gaps together with the
corresponding repair objectives, while $E_t$ records validated evidence about the
updated candidate.

The stages deliberately serve distinct purposes.
\emph{Diagnose} converts execution evidence into localized migration objectives;
\emph{Migrate} changes the candidate implementation; and
\emph{Verify} determines which properties have actually been established after the
change.
The Diagnoser and Verifier receive the candidate repository in read-only form,
represented by $\mathrm{ro}(\mathcal{J})$, while only the Migrator is allowed to modify
the repository.

This structure differs from open-ended iterative harness development.
The desired capability is already specified by the executable PyTorch repository, so
iteration is not used to discover an unknown objective.
Instead, it progressively reduces the behavioral discrepancy between the source
repository $\mathcal{P}$ and the migrated candidate $\mathcal{J}_t$.

Two persistent objects carry state across iterations.
The repository $\mathcal{J}_t$ records what the current implementation \emph{is},
whereas $E_t$ records validated knowledge about what remains incorrect or unestablished.
Keeping these objects separate prevents later iterations from having to reconstruct
development state solely from implementation code or interaction history.

Evidence produced after one migration step is subsequently available to diagnosis of the
next step, yielding the feedback flow
\begin{equation}
\mathcal{J}_{t-1}
\;\longrightarrow\;
D_t
\;\longrightarrow\;
\mathcal{J}_t
\;\longrightarrow\;
E_t
\;\longrightarrow\;
D_{t+1}.
\label{eq:appendix-harness-feedback-flow}
\end{equation}

Equation~(\ref{eq:appendix-harness-feedback-flow}) makes explicit that repository state
and validated diagnostic state are maintained separately while evidence is propagated
across iterations.

\subsection{Reference-Grounded Diagnostic Taxonomy}
\label{app:harness-diagnostics}

The diagnostic report
$\mathcal{G}(\mathcal{J};\mathcal{S},\mathcal{P})$
is constructed exclusively from the task specification, the task-visible PyTorch
repository, and the evolving JAX candidate.
Its purpose is not to assign a benchmark score, but to convert ordinary development-time
execution into localized evidence that can be consumed by the Diagnoser, Migrator, and
Verifier.

Concretely, each diagnostic records
(i) the property being examined,
(ii) the execution context used to examine it,
(iii) an outcome in
\{\emph{established}, \emph{violated}, \textsc{not established}\},
and
(iv) supporting observations such as runtime structure, relevant values or metadata,
and failure traces when execution does not complete.
This evidence is retained across iterations through $E_t$ and can therefore guide later
diagnosis without requiring the harness to infer previous failures solely from the
current repository state.

\paragraph{Development reference versus benchmark oracle.}
Under \methodname, benchmark reference outputs are recorded ahead of time from the
pure-PyTorch original and consumed only by the separate grading pipeline
(\S\ref{sec:bench-grading}).
Those frozen references never enter the development environment.

In contrast, \harnessname executes the task-visible PyTorch repository $\mathcal{P}$
itself during development and derives diagnostic observations directly from that
execution.
The PyTorch repository therefore acts as an \emph{executable specification}, not as a
source of benchmark goldens.
Development-time observations may be incomplete or inconclusive, and no frozen
reference, hidden test, verifier result, reference JAX solution, or benchmark reward is
exposed to the harness.
The complete information boundary is detailed in Appendix~\ref{app:harness-boundary}.

\paragraph{Differential behavioral grounding.}
The main reference-grounded diagnostic executes a matched probe against both
$\mathcal{P}$ and $\mathcal{J}$ through their native execution paths.
The probe constructs inputs through the repositories' exposed interfaces, performs the
corresponding computation, and records the resulting runtime contract.

The comparison primarily establishes whether the migrated implementation preserves
externally observable behavior that repository code depends on, including the structure
of returned objects and their exposed fields.
For example, the PyTorch implementation may return a structured model output while the
candidate exposes only its underlying array.
Although a candidate-authored numerical test could flatten both values and report
agreement, downstream repository code expecting the original interface would still
fail.
Differential execution exposes this discrepancy because the behavioral reference is
defined outside the candidate itself.

This distinction is important for avoiding circular self-validation.
Tests authored only against $\mathcal{J}$ can accidentally encode assumptions introduced
by the migration itself, whereas a matched execution against $\mathcal{P}$ anchors the
claim to behavior of the original repository.

\paragraph{Candidate-side diagnostics.}
Not every migration property requires differential execution.
A complementary set of diagnostics directly exercises properties of $\mathcal{J}$ that
follow from the task contract or from the semantics required by its JAX execution path.
These checks cover basic executability, independence from prohibited dependencies,
parameter representation, differentiability, compatibility with JAX transformations,
and integrated execution of the migrated training or serving entry point.

These diagnostics progressively narrow the source of a failure.
For example, failure to import the repository is distinguished from an invalid parameter
tree; successful construction is distinguished from a computation with no live gradient;
and local execution is distinguished from an integrated training or serving path that
fails only after components are composed.
The objective is localization rather than reproduction of the benchmark's grading
lattice.

\paragraph{Independence diagnostics.}
Independence is treated as a development-time property because it is already explicit in
the migration specification.
The candidate is exercised under conditions intended to expose reliance on dependencies
or external resources excluded by the task, including PyTorch-dependent execution and
network-dependent behavior.
Such checks can reveal dependencies that ordinary development accidentally masks, for
example through locally cached resources.

These checks instantiate only task-visible constraints.
They neither invoke nor approximate the hidden independence checks used by the benchmark
grader; the detailed separation between development and grading environments is given in
Appendix~\ref{app:harness-boundary}.

\paragraph{Progressive divergence localization.}
An end-to-end failure alone provides little information about where a cross-framework
migration diverged.
\harnessname therefore combines diagnostics at multiple levels of execution, moving from
repository validity and independence toward runtime contracts and integrated
training or serving behavior.

This organization shares the high-level intuition of the functional-equivalence lattice
in Appendix~\ref{app:lattice}: deeper execution can support stronger claims about the
candidate.
The two mechanisms are nevertheless separate.
The development diagnostics neither execute the hidden grader nor reconstruct its checks,
thresholds, references, or reward.
They are independently constructed from task-visible artifacts and are used only to
produce development-time evidence.

Table~\ref{tab:rgh-checks} summarizes the diagnostics used by \harnessname.
Differential checks use the executable PyTorch repository as an external behavioral
anchor, whereas candidate-side checks establish properties of the evolving JAX
repository directly.

\begin{table*}[t]
\centering
\scriptsize
\setlength{\tabcolsep}{4pt}
\caption{
Development-time diagnostics used by \harnessname.
\emph{Differential} checks compare observable behavior of the candidate with the
task-visible PyTorch repository; \emph{candidate} checks exercise properties implied by
the migration contract or the exposed execution path.
Returned evidence is used for localization and repair rather than benchmark scoring.
}
\vspace{0pt}
\label{tab:rgh-checks}
\begin{tabular}{
@{}
p{2.3cm}
p{1.7cm}
p{4.0cm}
p{5.0cm}
@{}}
\toprule
Check &
Grounding &
Property examined &
Evidence returned \\
\midrule

Module import
&
candidate
&
Required modules can be loaded through the migrated repository's intended import paths.
&
Import context together with successful construction or the exception that prevents the
repository from being initialized.
\\

Forbidden dependency
&
candidate
&
Execution does not rely on dependencies or external resources excluded by the migration
contract.
&
The dependency or resource access exposed during execution, or evidence that the tested
path executes without it.
\\

Output contract
&
differential
&
The migrated implementation preserves the externally visible runtime structure and
fields exposed by the PyTorch original.
&
Matched observations from $\mathcal{P}$ and $\mathcal{J}$, including output structure,
available fields, and any discrepancy between the two execution paths.
\\

Parameter tree
&
candidate
&
The migrated parameter representation forms a valid JAX pytree compatible with the
repository's execution path.
&
Observed tree structure and the construction or lookup failure associated with an
incompatible parameter representation.
\\

Gradient liveness
&
candidate
&
The training objective remains differentiable through the parameters intended to be
optimized.
&
Whether gradients can be produced for the relevant parameter path, together with the
failure context when differentiation is broken.
\\

JAX tracing
&
candidate
&
The relevant computation is compatible with the tracing assumptions required by the JAX
implementation.
&
Successful transformed execution or the trace/concretization failure that identifies the
incompatible computation.
\\

Training / serving entry
&
candidate
&
The repository's exposed end-to-end entry path executes coherently after migrated
components are composed.
&
Integrated execution status and the runtime trace at which the composed training or
serving path fails.
\\

\bottomrule
\end{tabular}
\vspace{0pt}
\end{table*}

\paragraph{Diagnostic outcome semantics.}
A diagnostic result describes what the corresponding probe actually establishes, rather
than treating every unsuccessful execution as evidence against the candidate.
For a property $q$, the report assigns one of three outcomes:

\begin{itemize}
    \item \emph{Established}: the diagnostic executed with sufficient evidence and
    observed behavior consistent with $q$.

    \item \emph{Violated}: the diagnostic executed with sufficient evidence and observed
    behavior inconsistent with $q$.

    \item \textsc{Not established}: the diagnostic itself could not reliably determine
    whether $q$ holds.
\end{itemize}

The distinction between \emph{violated} and \textsc{not established} is essential in
cross-framework migration.
Multiple correct JAX implementations can realize the same external behavior through
different internal structures or calling conventions.
A probe that assumes a convention not supported by a candidate may therefore fail before
it has produced evidence about the behavioral property it was intended to test.
Treating such a probe failure as a candidate failure would introduce false diagnostic
evidence.

\paragraph{Evidence preservation under uncertainty.}
A \textsc{not established} result is not discarded.
The harness retains the execution context and failure trace that made the diagnostic
inconclusive, allowing the next Diagnoser to decide whether the unresolved property
requires a different probe, a repository repair, or no immediate action.
Likewise, an \emph{established} result records the evidence supporting the property so
that later iterations need not repeatedly infer it from the repository.

Consequently, $E_t$ represents a partial and revisable account of the candidate rather
than a scalar notion of correctness.
The harness may know that some properties have been established, have concrete evidence
that others are violated, and remain uncertain about the rest.
This explicit representation of uncertainty is what allows execution evidence to be
carried forward through the diagnose--migrate--verify loop without conflating diagnostic
failure with migration failure.

\paragraph{Scope of a diagnostic claim.}
Development-time diagnostics are intentionally local.
Establishing one property does not certify functional equivalence of the repository as a
whole, and absence of a detected violation does not imply that the candidate will pass
the benchmark.
Only the separate, unchanged \methodname grading pipeline produces the final benchmark
measurement.

Accordingly, $\mathcal{G}(\mathcal{J};\mathcal{S},\mathcal{P})$ should be interpreted as
a collection of execution-grounded claims about properties that could be established
from task-visible information at the current iteration.
Its role is to improve localization and repair, not to serve as a surrogate benchmark
oracle.

\subsection{Detailed Information Boundary}
\label{app:harness-boundary}

Because \methodname explicitly guards against fabricated equivalence
(\S\ref{app:design-challenges}), a reference-grounded harness must make its information
boundary explicit.
\harnessname uses exactly the information granted to an ordinary agent by the task and
nothing from the benchmark's hidden grading state.
Table~\ref{tab:harness-boundary} summarizes this boundary.

\begin{table*}[t]
\centering
\small
\setlength{\tabcolsep}{8pt}
\caption{
Information boundary of \harnessname.
The harness uses only artifacts available to an ordinary agent during task execution
and never accesses hidden benchmark state.
}
\vspace{0pt}
\label{tab:harness-boundary}
\begin{tabular}{@{}p{6.7cm}p{6.7cm}@{}}
\toprule
\textbf{Agent-visible: used by \harnessname}
&
\textbf{Agent-hidden: never read}
\\
\midrule

\texttt{instruction.md};
\texttt{torch\_orig/};
the released checkpoint;
the candidate repository;
outputs and exceptions produced by executing task-visible artifacts
&
hidden tests;
frozen reference outputs;
the verifier implementation;
reference JAX solutions;
benchmark rewards, scores, and per-check grading results
\\

\bottomrule
\end{tabular}
\vspace{0pt}
\end{table*}

We enforce this boundary at both the filesystem and execution levels.
Development environments expose only task-visible components needed to solve the
migration rather than the hidden grading implementation.
Candidate state is exchanged through the evolving repository rather than through
verifier-generated grading artifacts.
Execution credentials and intermediate state are likewise restricted so that the
development environment cannot serve as an indirect route to hidden benchmark
information.

This distinction is important because information that is technically reachable can
affect the legitimacy of self-verification even when it does not appear in an observed
agent trace.
We therefore define the boundary according to what the harness can access, rather than
only according to what the agent happens to consume during one execution.

The same principle constrains the diagnostics themselves.
A diagnostic is included only when the property it examines can be derived from the task
specification, the task-visible PyTorch repository, or the ordinary semantics of the
exposed execution interface.
Diagnostics whose design depends on hidden verifier behavior are excluded from
\harnessname.

\paragraph{Unchanged final grading.}
\harnessname changes only the procedure used to construct the candidate repository.
Once migration terminates, the resulting repository is evaluated through the same
unmodified \methodname grading pipeline as every off-the-shelf baseline.
Neither the verifier nor the reward function is modified for \harnessname.

\subsection{Role Prompts for Diagnose--Migrate--Verify}
\label{app:harness-prompts}

For reproducibility, we provide the complete role prompts used by
\harnessname.
The prompts instantiate the three stages of the
\emph{diagnose--migrate--verify} loop described in
Appendix~\ref{app:harness-loop}.
They differ primarily in their responsibilities and permissions:
the \emph{Diagnoser} analyzes the current candidate and available execution evidence to
localize discrepancies and formulate repair objectives;
the \emph{Migrator} is the only role permitted to modify the evolving JAX repository;
and the \emph{Verifier} independently exercises the updated candidate and records which
properties are established, violated, or remain unresolved.

All three roles operate only on task-visible artifacts and development-time evidence
available to \harnessname.
The prompts do not expose hidden tests, frozen benchmark references, verifier internals,
reference JAX solutions, benchmark scores, or rewards.
Separating the roles in this way keeps diagnosis, repository modification, and
post-update verification explicit rather than conflating them within a single
open-ended coding-agent instruction.

The following listings give the exact prompts used by the three roles.

\promptlisting
{Diagnoser prompt.}
{lst:harness-diagnoser}
{sections/prompts/prompt_diagnoser.tex}

\promptlisting
{Migrator prompt.}
{lst:harness-migrator}
{sections/prompts/prompt_migrator.tex}

\promptlisting
{Verifier prompt.}
{lst:harness-verifier}
{sections/prompts/prompt_verifier.tex}

\end{document}